\documentclass[3p,times,twocolumn]{elsarticle}
\pdfoutput=1

\usepackage{ecrc}

\volume{00}

\firstpage{1}

\journalname{Nuclear Physics B Proceedings Supplement}

\runauth{}

\jid{nuphbp}

\jnltitlelogo{Nuclear Physics B Proceedings Supplement}

\usepackage{amssymb}

\usepackage{amsmath}

\usepackage[mathlines,pagewise]{lineno}

\biboptions{sort&compress}

\usepackage[figuresright]{rotating}

\allowdisplaybreaks

\begin{document}

\begin{frontmatter}



\dochead{}

\title{Update of the electroweak precision fit, interplay with 
  Higgs-boson signal strengths and model-independent constraints on
  new physics\tnoteref{tn}} 

\tnotetext[tn]{Based on a talk presented by S. Mishima in the 37th
  International Conference on High Energy Physics (ICHEP) held in
  Valencia, Spain on July 2-9, 2014.} 


\author[a]{Marco Ciuchini}
\author[b]{Enrico Franco}
\author[c,d]{Satoshi Mishima}
\author[e]{Maurizio Pierini}
\author[f]{Laura Reina}
\author[b]{Luca Silvestrini}

\address[a]{INFN, Sezione di Roma Tre, Via della Vasca Navale 84, I-00146 Roma, Italy}
\address[b]{INFN, Sezione di Roma, Piazzale A. Moro 2, I-00185 Roma, Italy}
\address[c]{Dipartimento di Fisica, Universit\`a di Roma ``La Sapienza'', Piazzale A. Moro 2, I-00185 Roma, Italy}
\address[d]{SISSA, Via Bonomea 265, I-34136 Trieste, Italy}
\address[e]{California Institute of Technology, 1200 E. California Blvd., Pasadena, CA 91125, USA}
\address[f]{Physics Department, Florida State University, Tallahassee, FL 32306-4350, USA}

\begin{abstract}
We present updated global fits of the Standard Model and beyond to
electroweak precision data, taking into account recent progress in
theoretical calculations and experimental measurements. From the 
fits, we derive model-independent constraints on new physics by
introducing oblique and epsilon parameters, and modified
$Zb\bar{b}$ and $HVV$ couplings. Furthermore,
we also perform fits of the scale factors of the Higgs-boson couplings
to observed signal strengths of the Higgs boson. 
\end{abstract}

\begin{keyword}
Electroweak precision fit \sep 
Higgs boson \sep 
Physics beyond the Standard Model 


\end{keyword}

\end{frontmatter}



\section{Introduction}
\label{sec:intro}

In 2012 a Higgs boson, possibly the last missing piece of the
Standard Model (SM), was discovered at the Large Hadron
Collider (LHC)~\cite{Aad:2012tfa,Chatrchyan:2012ufa}. The observed
properties of the discovered Higgs boson look very much like the SM ones. 
Furthermore, no new particle, except for the Higgs boson, has been
observed so far. Indirect searches for new physics (NP) are therefore
as relevant as ever after the LHC 7-8 TeV run. 

In this study we present a global electroweak (EW) precision fit which
provides severe constraints on any NP models relevant to solve the
hierarchy problem. Recent studies of the EW precision fit can also be
found, {\it e.g.}, in 
Refs.~\cite{Eberhardt:2012gv,Baak:2012kk,Erler:2012wz,Ciuchini:2013pca,Blas:2013ana,Baak:2014ora}. 
The precise measurements of the masses of the top quark, the Higgs
boson and the $W$ boson at the Tevatron and the LHC increase the
constraining power of  
the EW precision fit. On the theoretical side, full fermionic two-loop
EW corrections to the partial widths of the $Z$ boson have been
recently calculated~\cite{Freitas:2012sy,Freitas:2013dpa,Freitas:2014hra}. 
Consequently, theoretical uncertainties associated with missing
higher-order corrections are expected to be below the experimental
uncertainties~\cite{Freitas:2014owa}. For example, the current
theoretical and experimental uncertainties on the $W$-boson mass are 
4 MeV and 15 MeV, respectively. The theoretical uncertainties thus can 
be neglected in the fit at the current experimental precision. 
We perform a Bayesian analysis of the EW precision data in the SM and
beyond using the Bayesian Analysis Toolkit (BAT) 
library~\cite{Caldwell:2008fw}. 
We study NP contributions to EW precision observables (EWPO) in
a model-independent way by introducing oblique
parameters~\cite{Peskin:1990zt,Peskin:1991sw}, epsilon
parameters~\cite{Altarelli:1990zd,Altarelli:1991fk,Altarelli:1993sz}, 
modified $Zb\bar{b}$ couplings, and modified $HVV$ couplings to EW
vector bosons ($V$). 

\begin{table*}[t]
\centering
\begin{tabular}{lcccc}
\hline
& Data & Fit & Indirect & Pull \\
\hline
$\alpha_s(M_Z^2)$ &
  $0.1185\pm 0.0005$ & 
  $0.1185\pm 0.0005$ & 
  $0.1185\pm 0.0028$ & 
  $+0.0$ 
\\
$\Delta\alpha_{\rm had}^{(5)}(M_Z^2)$ &
  $0.02750\pm 0.00033$ & 
  $0.02741\pm 0.00026$ & 
  $0.02727\pm 0.00042$ & 
  $-0.4$ 
\\
$M_Z$ [GeV] &
  $91.1875\pm 0.0021$ & 
  $91.1879\pm 0.0020$ & 
  $91.198\pm 0.011$ & 
  $+0.9$ 
\\
$m_t$ [GeV] &
  $173.34\pm 0.76$ & 
  $173.6\pm 0.7$ & 
  $176.6\pm 2.5$ & 
  $+1.2$ 
\\
$m_H$ [GeV] &
  $125.5\pm 0.3$ & 
  $125.5\pm 0.3$ & 
  $99.9\pm 26.6$ & 
  $-0.8$ 
\\
\hline
$M_W$ [GeV] &
  $80.385\pm 0.015$ & 
  $80.367\pm 0.006$ & 
  $80.363\pm 0.007$ & 
  $-1.3$ 
\\
$\Gamma_W$ [GeV] &
  $2.085\pm 0.042$ & 
  $2.0892\pm 0.0005$ & 
  $2.0892\pm 0.0005$ & 
  $+0.1$ 
\\
$\Gamma_{Z}$ [GeV] &
  $2.4952\pm 0.0023$ & 
  $2.4945\pm 0.0004$ & 
  $2.4944\pm 0.0004$ & 
  $-0.3$ 
\\
$\sigma_{h}^{0}$ [nb] &
  $41.540\pm 0.037$ & 
  $41.488\pm 0.003$ & 
  $41.488\pm 0.003$ & 
  $-1.4$ 
\\
$\sin^2\theta_{\rm eff}^{\rm lept}(Q_{\rm FB}^{\rm had})$ &
  $0.2324\pm 0.0012$ & 
  $0.23145\pm 0.00009$ & 
  $0.23144\pm 0.00009$ & 
  $-0.8$ 
\\
$P_\tau^{\rm pol}$ &
  $0.1465\pm 0.0033$ & 
  $0.1476\pm 0.0007$ & 
  $0.1477\pm 0.0007$ & 
  $+0.3$ 
\\
$\mathcal{A}_\ell$ (SLD) &
  $0.1513\pm 0.0021$ & 
  $0.1476\pm 0.0007$ & 
  $0.1471\pm 0.0007$ & 
  $-1.9$ 
\\
$\mathcal{A}_{c}$ &
  $0.670\pm 0.027$ & 
  $0.6682\pm 0.0003$ & 
  $0.6682\pm 0.0003$ & 
  $-0.1$ 
\\
$\mathcal{A}_{b}$ &
  $0.923\pm 0.020$ & 
  $0.93466\pm 0.00006$ & 
  $0.93466\pm 0.00006$ & 
  $+0.6$ 
\\
$A_{\rm FB}^{0,\ell}$ &
  $0.0171\pm 0.0010$ & 
  $0.0163\pm 0.0002$ & 
  $0.0163\pm 0.0002$ & 
  $-0.8$ 
\\
$A_{\rm FB}^{0,c}$ &
  $0.0707\pm 0.0035$ & 
  $0.0740\pm 0.0004$ & 
  $0.0740\pm 0.0004$ & 
  $+0.9$ 
\\
$A_{\rm FB}^{0,b}$ &
  $0.0992\pm 0.0016$ & 
  $0.1035\pm 0.0005$ & 
  $0.1039\pm 0.0005$ & 
  $+2.8$ 
\\
$R^{0}_{\ell}$ &
  $20.767\pm 0.025$ & 
  $20.752\pm 0.003$ & 
  $20.752\pm 0.003$ & 
  $-0.6$ 
\\
$R^{0}_{c}$ &
  $0.1721\pm 0.0030$ & 
  $0.17224\pm 0.00001$ & 
  $0.17224\pm 0.00001$ & 
  $+0.0$ 
\\
$R^{0}_{b}$ &
  $0.21629\pm 0.00066$ & 
  $0.21578\pm 0.00003$ & 
  $0.21578\pm 0.00003$ & 
  $-0.8$ 
\\
\hline
\end{tabular}
\caption{Experimental data and SM fit results for the five input
  parameters and fifteen EWPO considered in this study. The values in
  the column ``Indirect'' are determined without using the
  corresponding experimental information, while those in the column
  ``Pull'' represent the pulls in units of standard
  deviations~\cite{Bona:2005vz}.} 
\label{tab:SMfit}
\end{table*}

Moreover we also derive constraints on Higgs-boson couplings from the
experimental data on the signal strengths of the Higgs boson measured
at the Tevatron and the LHC. We consider only the couplings which have the
same tensor structures as in the SM, and introduce the scale factors
$\kappa_V$ and $\kappa_f$ for the $HVV$ and $Hf\bar{f}$ couplings to
SM vector bosons and fermions ($f$), respectively. 
Constraints from the EWPO are also investigated.  

The paper is organized as follows: 
in Sec.~\ref{sec:SM} we present our implementation of the EW precision
fit of the SM in some detail. 
Model-independent constraints on NP from the EW precision
fits are studied in Sec.~\ref{sec:NP}. 
In Sec.~\ref{sec:kappa} we derive constraints on the Higgs-boson
couplings from the data on the Higgs-boson signal strengths and the
EWPO. Finally we give a brief summary in Sec.~\ref{sec:summary}.

\section{Electroweak precision fit in the Standard Model}
\label{sec:SM}

Here we present the EW precision fit of the SM that we have performed
in our analysis. Details on the EWPO considered in the fit can be
found in Ref.~\cite{Ciuchini:2013pca} and references therein.  
Compared to our previous analysis in
Ref.~\cite{Ciuchini:2013pca} by four of the current authors, 
we update the 
data\footnote{The inclusion of the recent measurements of the effective
weak mixing angle at the hadron 
colliders~\cite{Chatrchyan:2011ya,TheATLAScollaboration:2013bha,Aaltonen:2013wcp,Aaltonen:2014loa,Abazov:2014jti}
does not alter our fit results significantly.}
on the strong coupling constant $\alpha_s(M_Z^2)$~\cite{Beringer:1900zz}, 
the top-quark mass $m_t$~\cite{ATLAS:2014wva},\footnote{Recent data
  from CMS~\cite{CMS-PAS-TOP-14-001} and D0~\cite{Abazov:2014dpa} as
  well as the Tevatron combination in Ref.~\cite{Tevatron:2014cka} are
  not considered here.} 
and the Higgs-boson mass
$m_H$~\cite{Aad:2014aba,CMS-PAS-HIG-13-001,Chatrchyan:2013mxa}, 
and use the recent theoretical expressions for the observables related
to the $Z$-boson partial
widths~\cite{Freitas:2012sy,Freitas:2013dpa,Freitas:2014hra}, 
which include the full fermionic two-loop EW contributions. 

The pole mass of the top quark reported by the hadron-collider
experiments is subject to ambiguities due to the renormalon
contribution and to the modeling of parton showers, colour
reconnection, and other technical details of the Monte Carlo (MC)
programs used in experimental 
analyses~\cite{Skands:2007zg,Wicke:2008iz,Buckley:2011ms}. 
It is believed that the ambiguity is at the level of 250 to 
500 MeV~\cite{ManganoTOP2013},\footnote{In
  Refs.~\cite{Moch:2014tta,Moch:2014lka}, 
the MC mass is converted into the pole mass via a short-distance
mass at a low scale, and the difference between the MC and pole masses
is estimated to be of the order of 1 GeV.} 
which does not affect significantly the EW precision fit at the
current experimental precision. Hence we do not consider it in the
current analysis. 

In Table~\ref{tab:SMfit} we present the results of the fit to the 
EWPO considered in our analysis together with the corresponding
experimental measurements (data). In the fourth column, we also
present the indirect determinations of the input parameters and 
the EWPO, obtained by assuming a flat prior for the
parameter or the observable under consideration. The values in the
last column show the compatibility between the data and the indirect
determination~\cite{Bona:2005vz}. We observe sizable deviations from
the SM in $\mathcal{A}_\ell$ and $A_{\mathrm{FB}}^{0,b}$ by 
$-1.9\sigma$ and $2.8\sigma$, respectively.\footnote{Adopting 
$\Delta\alpha_{\rm had}^{(5)}(M_Z^2)=0.02757\pm
0.00010$~\cite{Davier:2010nc} instead of the value in Table~\ref{tab:SMfit}, 
the pull values for $\mathcal{A}_\ell$ and $A_{\mathrm{FB}}^{0,b}$ 
become $-2.0\sigma$ and $2.5\sigma$, respectively, which are in
agreement with those in Ref.~\cite{Baak:2014ora}.}

\section{Model-independent constraints on new physics from the
  electroweak precision data}
\label{sec:NP}

In this section, we fit NP parameters to the EWPO, together with the
five SM parameters in Table~\ref{tab:SMfit}. The fit results for the
SM parameters will not be presented below, since they are similar to
the ones in the SM.

\begin{figure}[t!]
  \centering
  \vspace{-1mm}
  \includegraphics[width=.24\textwidth]{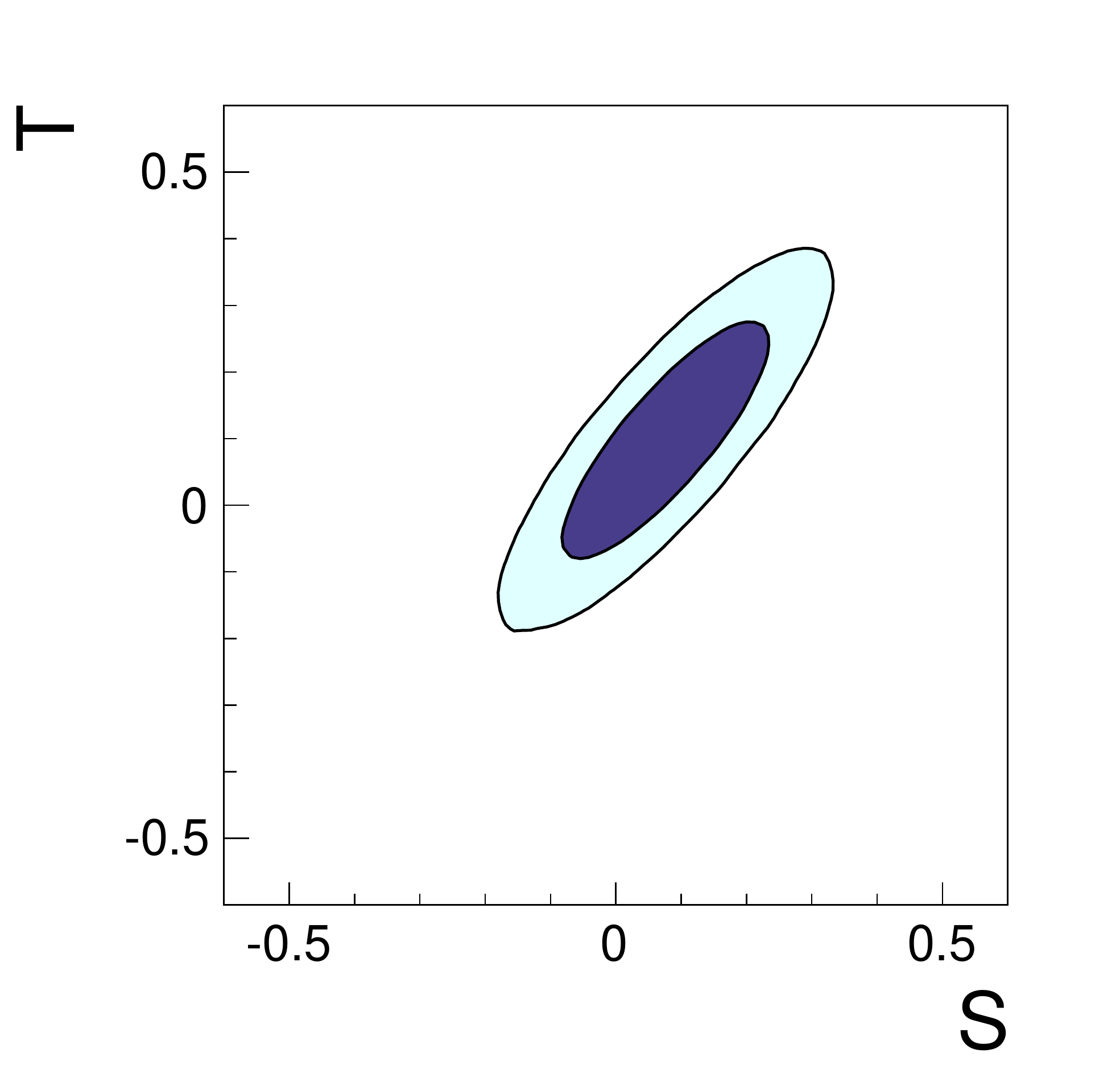}
  \hspace{-3mm}
  \includegraphics[width=.24\textwidth]{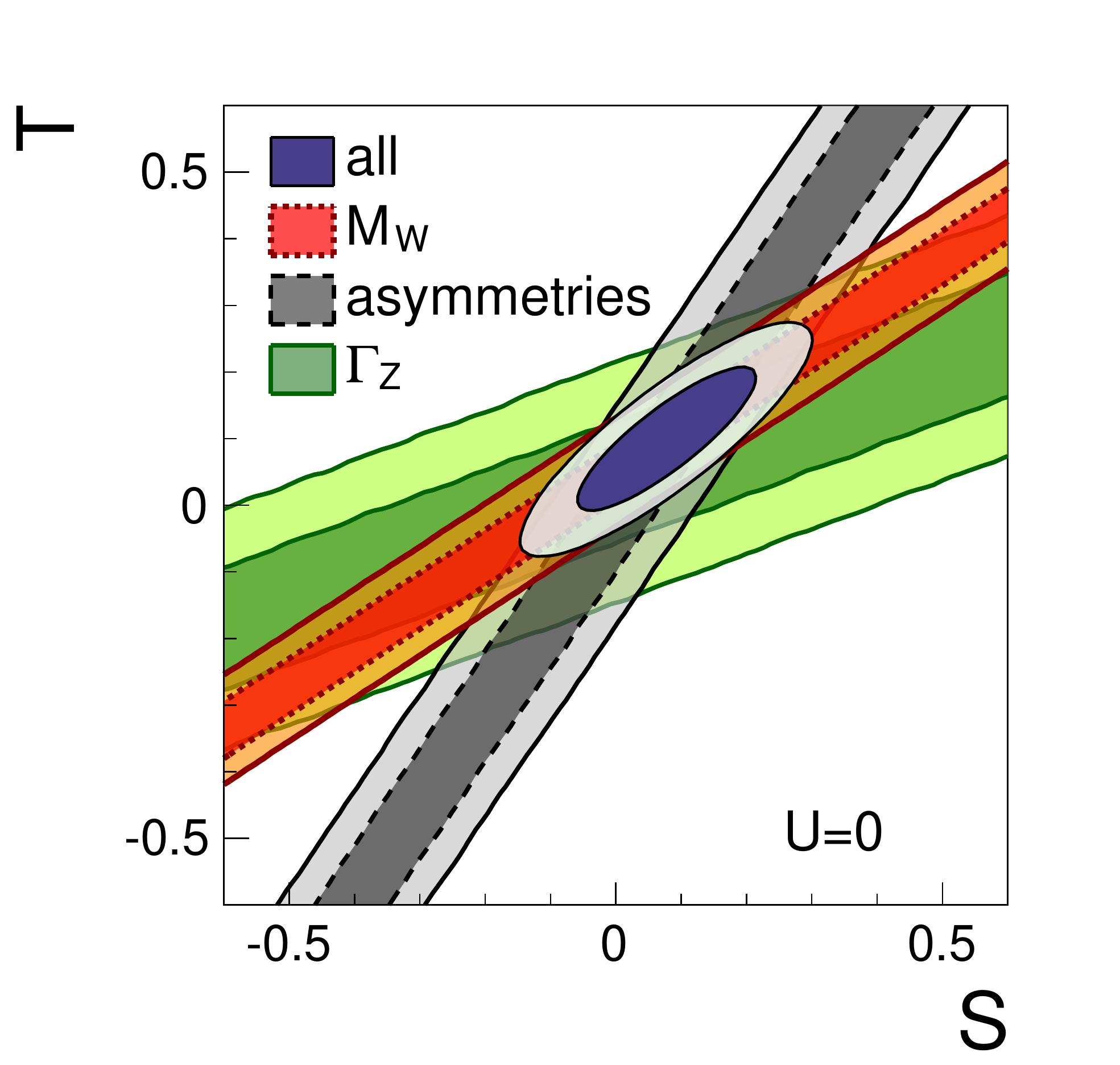}

  \vspace{-2mm}
  \caption{Two-dimensional probability distributions for the oblique
    parameters $S$ and $T$ at $68\%$ (the dark region) and $95\%$ (the
    light region) floating $U$ (left) or fixing $U=0$ (right). 
    In the right plot, the individual constraints from $M_W$, the
    asymmetry parameters $\sin^2\theta_{\rm eff}^{\rm lept}$,
    $P_\tau^{\rm pol}$, $A_f$ and $A_{\rm FB}^{0,f}$ with
    $f=\ell,c,b$, and $\Gamma_Z$ are also presented.} 
  \label{fig:Oblique}
\end{figure}

First we present fit results for the oblique parameters $S$, $T$, and
$U$ introduced in Ref.~\cite{Peskin:1990zt,Peskin:1991sw}. Those
parameters are useful for models where dominant NP effects appear
in the vacuum-polarization amplitudes of the EW gauge bosons. When the EW
symmetry is realized linearly, the parameter $U$ is associated with a
dimension-eight operator, and thus smaller than the others. The EWPO
considered in the current study depend on the three combinations of
the oblique parameters introduced in Ref.~\cite{Ciuchini:2013pca}.  
We summarize our fit results in Tables~\ref{tab:STUfit} and \ref{tab:STfit}
and in Fig.~\ref{fig:Oblique}. They do not show evidence for NP and are
in agreement with those reported in Refs.~\cite{Erler:2012wz,Baak:2014ora}. 

\begin{table}[t]
\centering
\begin{tabular}{c|c|rrr}
\hline
& Fit result & 
\multicolumn{3}{|c}{Correlations}
\\
\hline
$S$
& $0.08\pm 0.10$ 
& $1.00$ 
\\
$T$
& $0.10\pm 0.12$ 
& $0.85$ & $1.00$ 
\\
$U$
& $0.00\pm 0.09$ 
& $-0.49$ & $-0.79$ & $1.00$ 
\\
\hline
\end{tabular}
\caption{Fit results for the oblique parameters.}
\label{tab:STUfit}
\end{table}

\begin{table}[t]
\centering
\begin{tabular}{c|c|rrr}
\hline
& Fit result & 
\multicolumn{2}{|c}{Correlations}
\\
\hline
$S$
& $0.08\pm 0.09$ 
& $1.00$ 
\\
$T$
& $0.10\pm 0.07$
& $0.87$ & $1.00$ 
\\
\hline
\end{tabular}
\caption{Fit results for the oblique parameters fixing $U=0$.}
\label{tab:STfit}
\end{table}

\begin{figure}[t!]
  \centering
  \vspace{-1mm}
  \includegraphics[width=.24\textwidth]{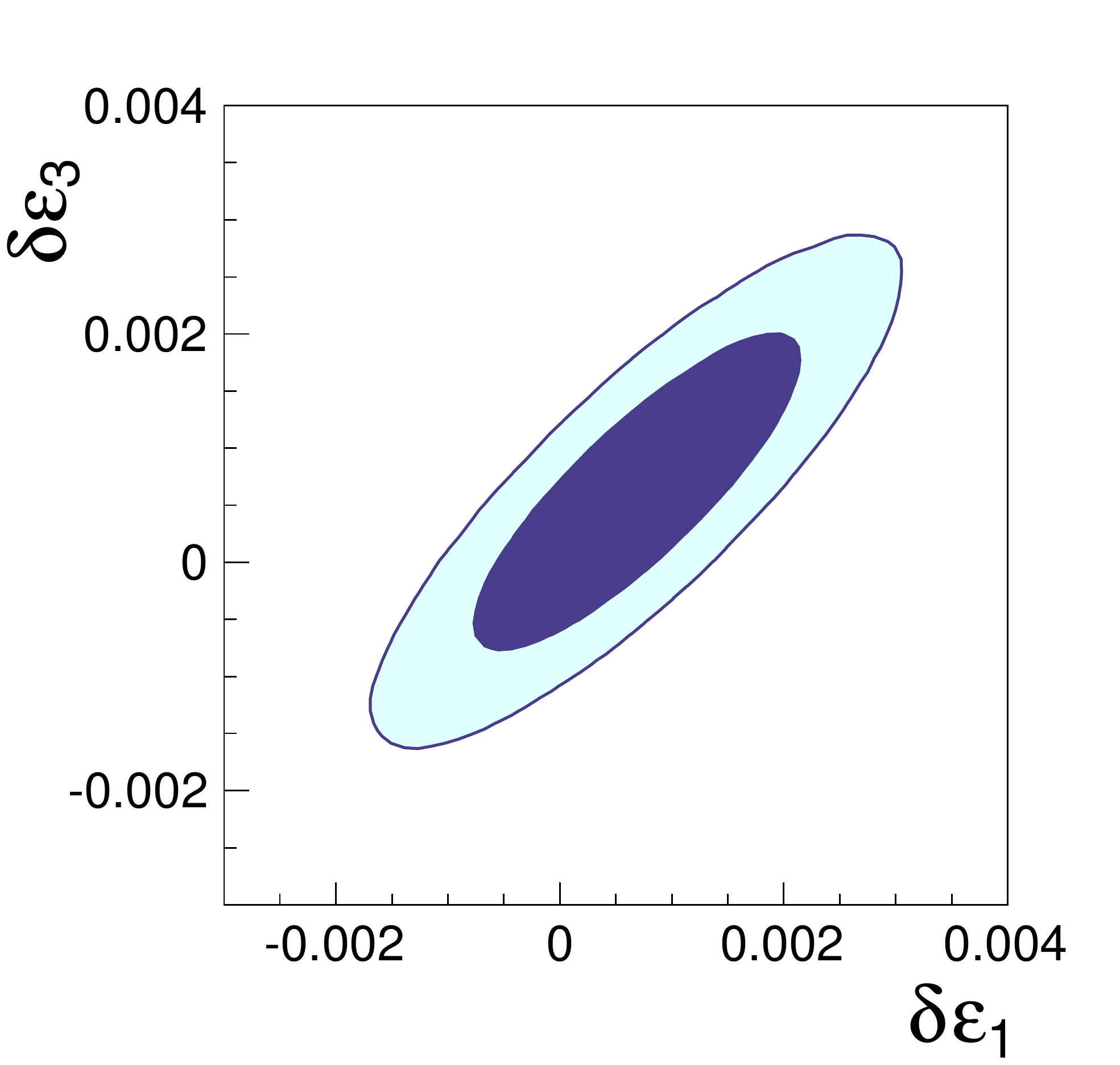}
  \hspace{-3mm}
  \includegraphics[width=.24\textwidth]{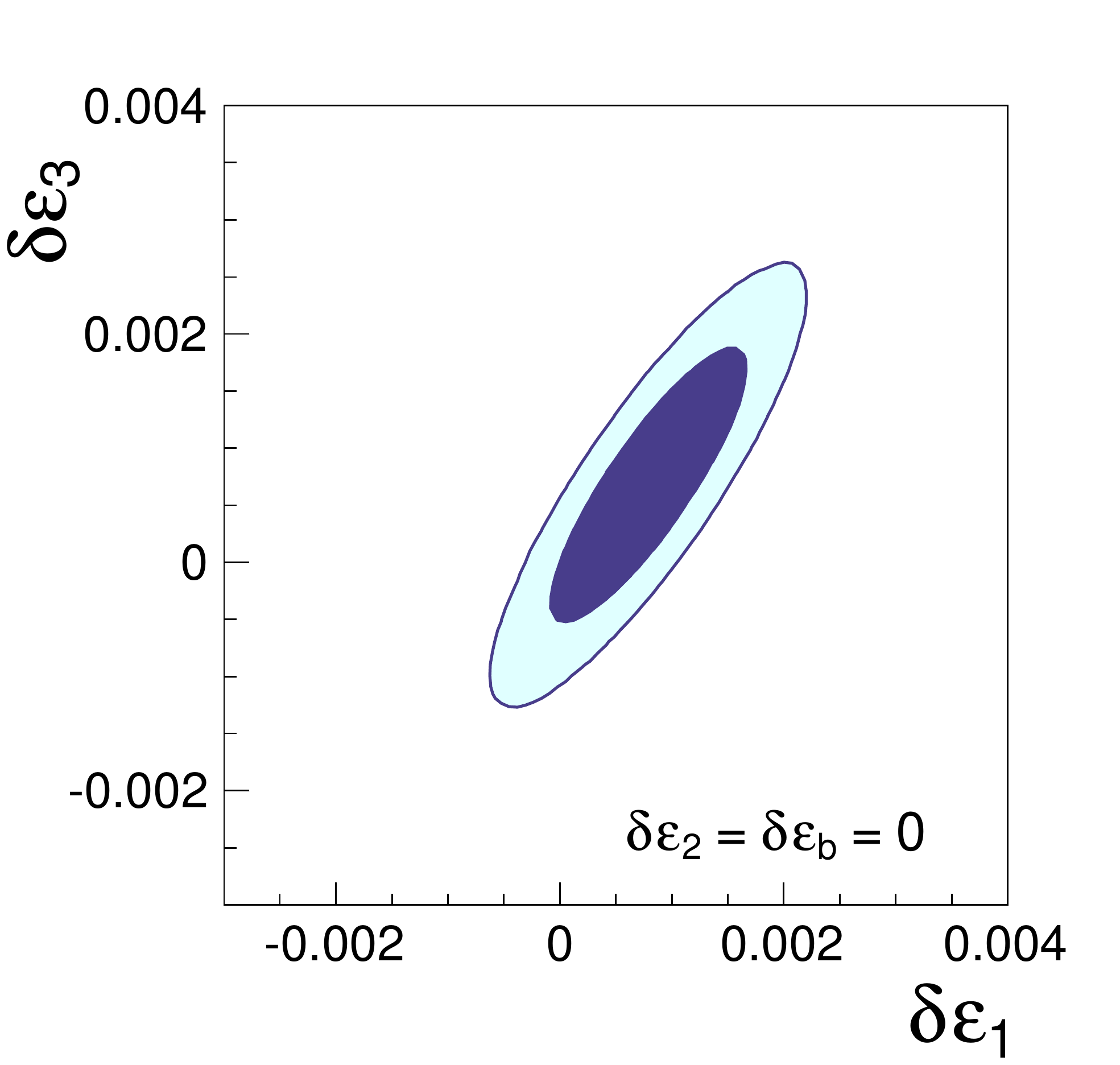}

  \vspace{-2mm}
  \caption{Two-dimensional probability distributions for 
    $\delta\epsilon_1$ and $\delta\epsilon_3$  
    at $68\%$ (the dark region) and $95\%$ (the light region)
    floating all $\delta\epsilon_i$ parameters (left) or
    fixing $\delta\epsilon_2=\delta\epsilon_b=0$ (right).}
  \label{fig:deps}
\end{figure}

Next we consider the epsilon parameters introduced in
Refs~\cite{Altarelli:1990zd,Altarelli:1991fk,Altarelli:1993sz}.  
Unlike the $S$, $T$, and $U$ parameters discussed above, the epsilon
parameters involve SM contributions associated with the top quark and
the Higgs boson. Moreover, they also involve flavour non-universal
vertex corrections in the SM~\cite{Ciuchini:2013pca} and the
vacuum-polarization corrections that are not taken into account in the 
$S$, $T$, and $U$ parameters~\cite{Barbieri:2004qk}. Since all the SM
parameters, including $m_t$ and $m_H$, have now been measured, we
separate the NP contribution from the SM one, by defining,
\begin{linenomath*}
\begin{equation}
\delta\epsilon_i=\epsilon_i - \epsilon_{i,\mathrm{SM}}\qquad
\mathrm{for}\ \ i=1,2,3,b,
\end{equation}
\end{linenomath*}
where $\epsilon_i$ are the original epsilon parameters. Here and in
the following, a quantity with the subscript ``SM'' represents the
corresponding SM contribution. Using $\delta\epsilon_i$, the $W$-boson
mass and the effective vector and axial-vector couplings for the
$Zf\bar{f}$ interactions are given by 
\begin{linenomath*}
\begin{align}
M_W &=
M_{W,\mathrm{SM}}
\Bigg\{
1 
- \frac{1}{2(c_{W,\mathrm{SM}}^2 - s_{W,\mathrm{SM}}^2)}
\nonumber\\
&\hspace{10mm}\times
\Big[
- c_0^2 \delta\epsilon_1
+ (c_0^2-s_0^2) \delta\epsilon_2
+ 2s_0^2 \delta\epsilon_3
\Big]
\Bigg\}\,,
\\
g_V^f &=
g_{V,\mathrm{SM}}^f
+ 
\Big(
g_{V,\mathrm{SM}}^f - g_{A,\mathrm{SM}}^f
\Big) 
\Bigg(
\frac{\delta\epsilon_3-c_0^2\,\delta\epsilon_1}
{c_0^2 - s_0^2}
-
\delta\epsilon_b
\Bigg)
\nonumber\\
&\hspace{5mm}
+ 
\frac{g_{V,\mathrm{SM}}^f}{2}\,
\big(\delta \epsilon_1 + 2\delta \epsilon_b\big)\,,
\\
g_A^f 
&=
g_{A,\mathrm{SM}}^f 
+ \frac{I_3^f}{2}
\big( \delta \epsilon_1 + 2\delta \epsilon_b \big)\,,
\end{align}
\end{linenomath*}
where $\delta\epsilon_b=0$ for $f\neq b$, 
$I_3^f$ is the third component of weak isospin of fermion $f$, 
and $s_W$, $c_W$, $s_0$ and $c_0$ are defined in
Ref.~\cite{Ciuchini:2013pca}. Using the above effective couplings, the
$Z$-pole observables are calculated with the formulae presented in
Appendix A of Ref.~\cite{Ciuchini:2013pca}. Our fit results for the
$\delta\epsilon_i$ parameters are summarized in Tables~\ref{tab:4deps}
and \ref{tab:2deps}, where $\delta\epsilon_2$ and $\delta\epsilon_b$
are set to be zero in the latter. The corresponding two-dimensional
probability distributions for $\delta\epsilon_1$ and
$\delta\epsilon_3$ are plotted in Fig.~\ref{fig:deps}. 
The results are consistent with the SM.  

\begin{table}[t]
\setlength{\tabcolsep}{4pt}
\centering
\begin{tabular}{c|c|rrrr}
\hline
& Fit result & 
\multicolumn{4}{|c}{Correlations}
\\
\hline
$\delta\epsilon_1$ 
& $\phantom{-}0.0007 \pm 0.0010$
& $1.00$ 
\\
$\delta\epsilon_2$ 
& $-0.0001 \pm 0.0009$
& $0.80$ & $1.00$ 
\\
$\delta\epsilon_3$
& $\phantom{-}0.0006 \pm 0.0009$
& $0.86$ & $0.51$ & $1.00$ 
\\
$\delta\epsilon_b$
& $\phantom{-}0.0003 \pm 0.0013$
& $-0.33$ & $-0.32$ & $-0.22$ & $1.00$
\\
\hline
\end{tabular}
\caption{Fit results for the $\delta\epsilon_i$ parameters.}
\label{tab:4deps}
\end{table}

\begin{table}[t!]
\centering
\begin{tabular}{c|c|rr}
\hline
& Fit result & 
\multicolumn{2}{|c}{Correlations}
\\
\hline
$\delta\epsilon_1$
& $0.0008\pm 0.0006$
& $1.00$ 
\\
$\delta\epsilon_3$
& $0.0007\pm 0.0008$
& $0.87$ & $1.00$
\\
\hline
\end{tabular}
\caption{Fit results for $\delta\epsilon_1$ and $\delta\epsilon_3$ 
fixing $\delta\epsilon_2=\delta\epsilon_b=0$.}
\label{tab:2deps}
\end{table}

\begin{table}[t]
\centering
\begin{tabular}{c|c|rr}
\hline
& Fit result
& \multicolumn{2}{|c}{Correlations}
\\
\hline
$\delta g_R^b$ 
& $0.018\pm 0.007$
& $1.00$
\\
$\delta g_L^b$ 
& $0.0029\pm 0.0014$
& $0.90$ & $1.00$
\\
\hline
$\delta g_V^b$ 
& $\phantom{-}0.021\pm 0.008$
& $1.00$
\\
$\delta g_A^b$
& $-0.015\pm 0.006$
& $-0.98$ & $1.00$
\\
\hline
\end{tabular}
\caption{Fit results for the shifts in the $Zb\bar b$ couplings.}
\label{tab:Zbb}
\end{table}

\begin{figure}[t]
  \centering
  \vspace{-1mm}
  \includegraphics[width=.24\textwidth]{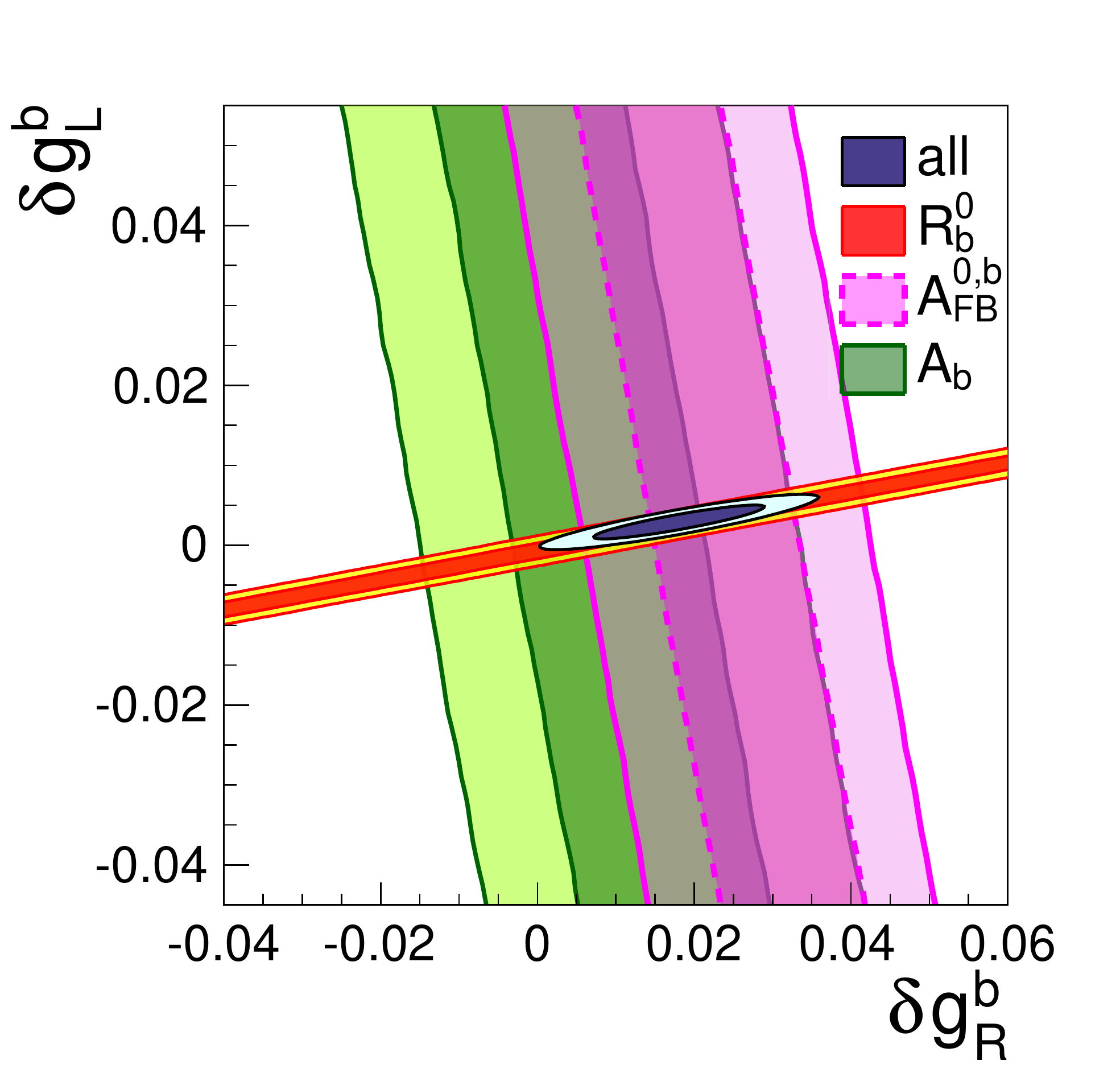}
  \hspace{-3mm}
  \includegraphics[width=.24\textwidth]{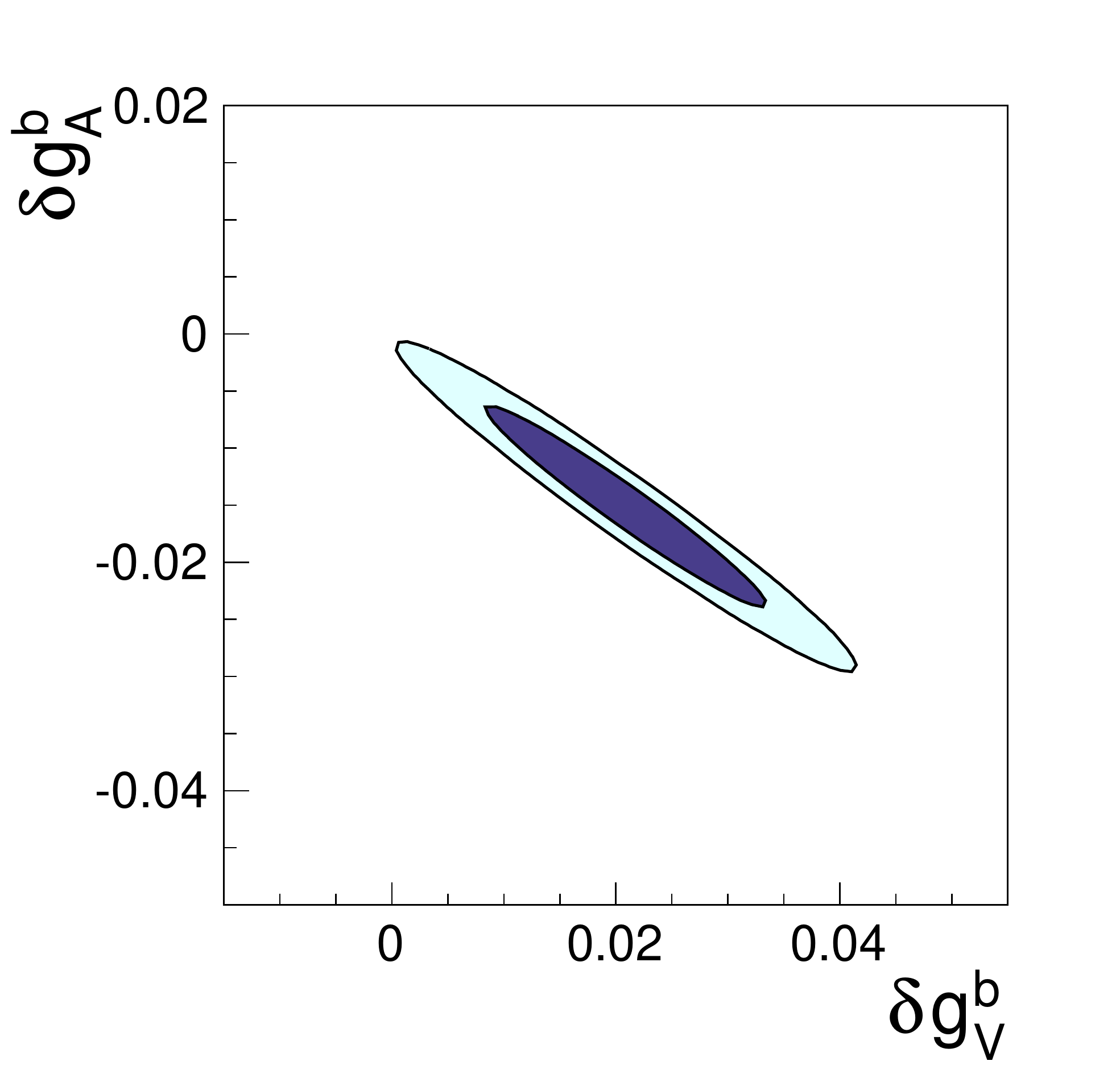}

  \vspace{-2mm}
  \caption{Two-dimensional probability distributions for 
    $\delta g_R^b$ and $\delta g_L^b$ (left), or
    $\delta g_V^b$ and $\delta g_A^b$ (right) 
    at $68\%$ (the dark region) and $95\%$ (the light region). 
    The individual constraints in the left plot are computed by
    omitting $\mathcal{A}_b$, $A_{\rm FB}^{0,b}$, $\Gamma_Z$,
    $\sigma_h^0$, $R_\ell^0$, $R_c^0$ and $R_b^0$ except when
    specified in the legend.} 
  \label{fig:Zbb}
\end{figure}

We also consider the case where dominant NP contributions appear in
the $Zb\bar{b}$ couplings (see, {\it e.g.}, Ref.~\cite{Batell:2012ca}
and references therein). We parameterize NP contributions to the
$Zb\bar{b}$ couplings as follows: 
\begin{linenomath*}
\begin{align}
g_i^b 
&= g^b_{i,\mathrm{SM}} + \delta g_i^b\qquad
\mathrm{for}\ \ i=R,L,V,A,
\end{align}
\end{linenomath*}
where the definitions of these couplings are given in
Ref.~\cite{Ciuchini:2013pca}. The EW precision fit provides four
solutions for the couplings, while two of them are disfavored by the
off $Z$-pole data for the forward-backward asymmetry in 
$e^+e^-\to b\bar{b}$~\cite{Choudhury:2001hs}. 
In Tables~\ref{tab:Zbb} and Fig.~\ref{fig:Zbb}, we present the
solution that is closer to the SM. We observe significant deviations
from the SM, which are attributed to the measured value of 
$A_{\mathrm{FB}}^{0,b}$.

\begin{figure}[t!]
  \centering
  \vspace{-1mm}
  \includegraphics[width=.24\textwidth]{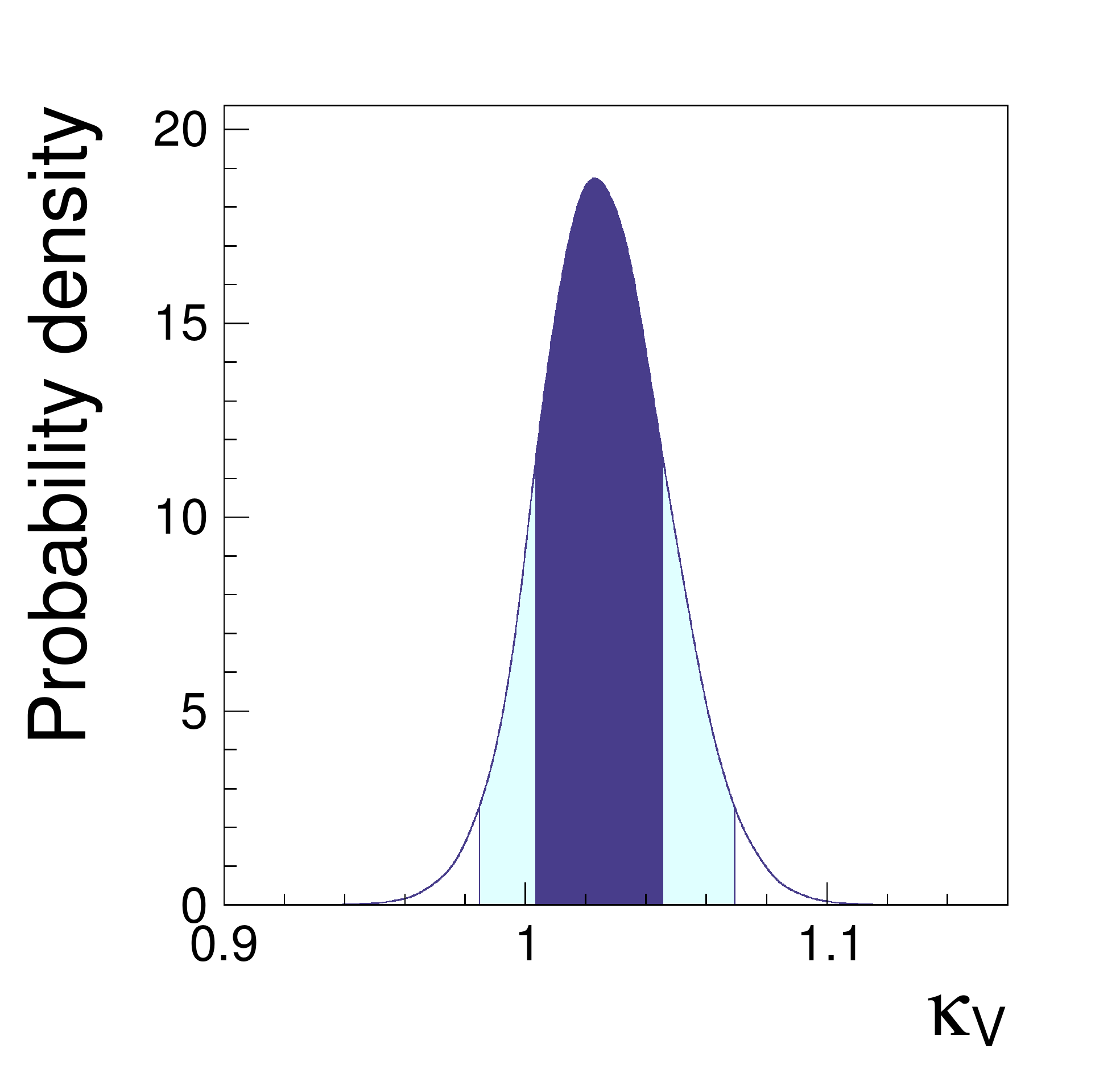}
  \hspace{-3mm}
  \includegraphics[width=.24\textwidth]{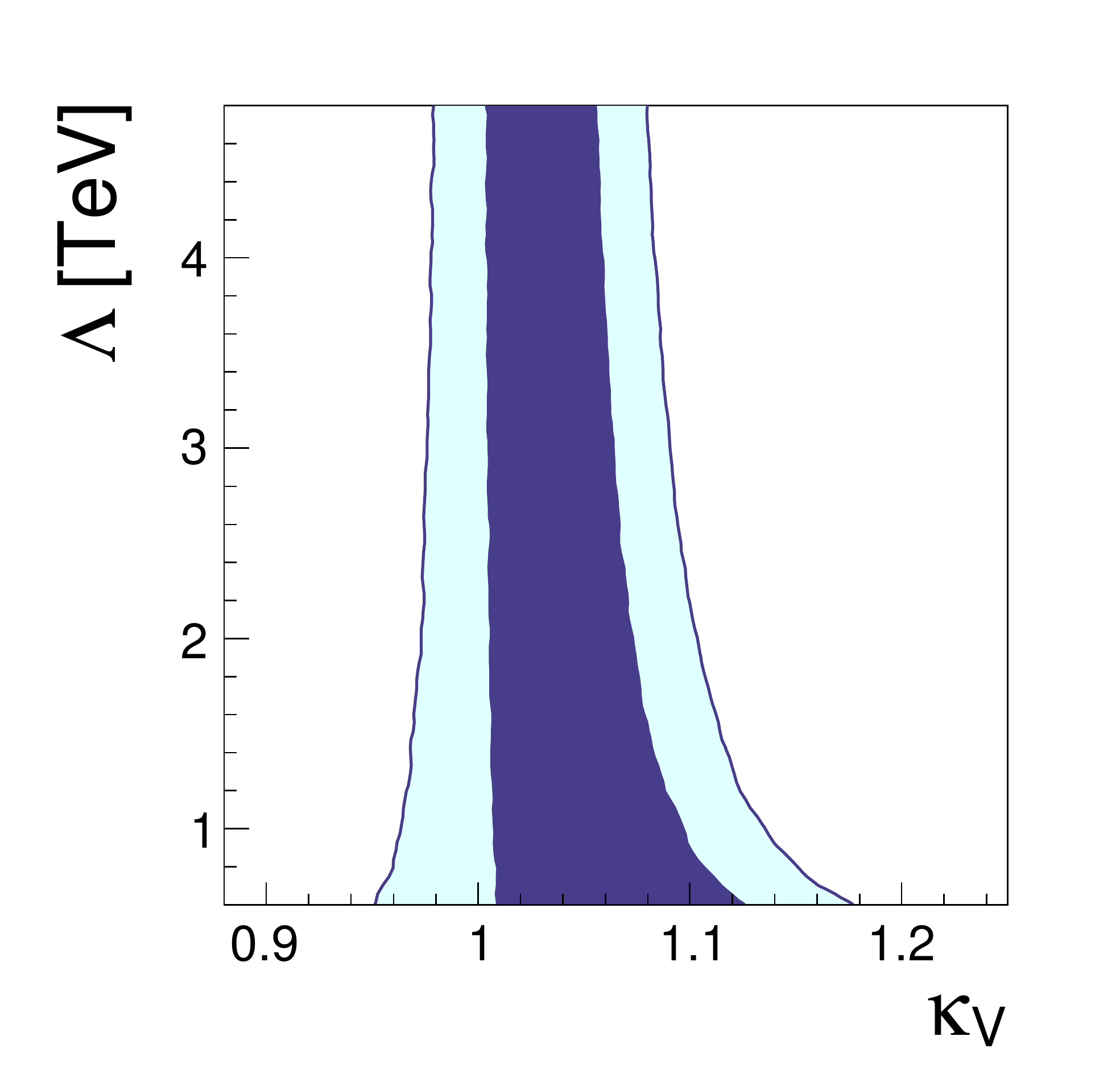}

  \vspace{-2mm}
  \caption{Left: Probability distribution for the scale factor 
    $\kappa_V$. Right: Two-dimensional probability distributions for  
    $\kappa_V$ and $\Lambda$. 
    The dark and light regions correspond to $68\%$ and $95\%$
    probabilities, respectively.}
  \label{fig:HVV}
\end{figure}

In various NP models the Higgs-boson couplings to the SM vector bosons and
fermions deviate from their SM values. It is therefore of interest to
study constraints on the Higgs-boson couplings from the EW precision test. 
We consider a general effective Lagrangian for a light Higgs-boson-like
scalar field $H$, assuming an approximate custodial symmetry and 
no other new light states below a cutoff 
scale~\cite{Giudice:2007fh,Contino:2010mh,Azatov:2012bz,Contino:2013kra}: 
\begin{linenomath*}
\begin{equation}
\mathcal{L} =
\frac{v^2}{4}{\rm Tr}\big(D_\mu\Sigma^\dagger D^\mu\Sigma\big)
\left( 1 + 2\kappa_V\frac{H}{v} + \cdots \right)
+ \cdots,
\label{eq:L_LightHiggs}
\end{equation}
\end{linenomath*}
where $v$ is the vacuum expectation value of the Higgs-boson field, 
and the longitudinal components of the $W$ and $Z$ bosons,
$\chi^a(x)$, are described by the two-by-two matrix  
$\Sigma(x) = \exp(i\tau^a\chi^a(x)/v)$ with $\tau^a$ being the Pauli
matrices. The deviation in the $HVV$ couplings is parameterized by 
the scale factor $\kappa_V$, which is equal to one in the SM. The
oblique parameters $S$ and $T$ then receive the following
contributions~\cite{Barbieri:2007bh}: 
\begin{linenomath*}
\begin{align}
S &= \frac{1}{12\pi} (1 - \kappa_V^2)
  \ln\bigg(\frac{\Lambda^2}{m_H^2}\bigg)\,,
\label{eq:S}
\\
T &= - \frac{3}{16\pi c_W^2} (1 - \kappa_V^2)
  \ln\bigg(\frac{\Lambda^2}{m_H^2}\bigg)\,,
\label{eq:T}
\end{align}
\end{linenomath*}
where $\Lambda = 4\pi v/\sqrt{|1-\kappa_V^2|}$ is the cutoff scale of
the effective Lagrangian. We present fit results for $\kappa_V$ in
Table~\ref{tab:HVV} and Fig.~\ref{fig:HVV}. Typical NP models, such as
composite Higgs models, generate smaller $\kappa_V$ ($\kappa_V<1$),
while larger $\kappa_V$ requires that the $W_LW_L$ scattering is
dominated by an isospin-two 
channel~\cite{Falkowski:2012vh,Bellazzini:2014waa}. The present
fit disfavors smaller $\kappa_V$, where the lower bound at 95\%
corresponds to the cutoff scale $\Lambda=18$ TeV. In the right plot of
Fig.~\ref{fig:HVV}, we generalize the analysis allowing for 
$\Lambda < 4\pi v/\sqrt{|1-\kappa_V^2|}$ and assuming that the
dynamics at the cutoff does not contribute sizably to the oblique
parameters. We find that $\kappa_V$ is tightly
constrained for $\Lambda>1$ TeV. Extra contributions to the oblique
parameters were studied, {\it e.g.}, in 
Refs.~\cite{Grojean:2006nn,Azatov:2013ura,Pich:2012dv,Pich:2013fea}. 

\begin{table}[t!]
\centering
\begin{tabular}{c|cc}
\hline
& 68\% & 95\%
\\
\hline
$\kappa_V$ & $1.025\pm 0.021$ & $[0.985,\, 1.069]$
\\
\hline
\end{tabular}
\caption{Fit results for the scale factor $\kappa_V$ at 68\% and
  95\% probabilities.} 
\label{tab:HVV}
\end{table}

\section{Constraints on the Higgs-boson couplings from the Higgs-boson
  and electroweak precision data}
\label{sec:kappa}

In this section we fit the Higgs-boson couplings to the data for the
Higgs-boson signal strengths and the EWPO, where the former are taken
from Refs.~\cite{CMS-PAS-HIG-13-001,ATLAS-CONF-2013-012}
for $H \to \gamma \gamma$,
Refs.~\cite{ATLAS-CONF-2013-013,CMS-PAS-HIG-13-002} for $H \to Z Z$,
Refs.~\cite{ATLAS-CONF-2013-030,Chatrchyan:2013iaa} for $H \to W^+ W^-$, 
Refs.~\cite{ATLAS-CONF-2013-108,CMS-PAS-HIG-13-004} for $H \to\tau^+\tau^-$, 
and
Refs.~\cite{Group:2012zca,ATLAS-CONF-2013-079,ATLAS-CONF-2014-011,Chatrchyan:2013zna,CMS-PAS-HIG-13-019}
for $H\to b\bar{b}$ (see also Ref.~\cite{Bechtle:2014ewa}). 
We consider the scale factors $\kappa_V$ and $\kappa_f$ for the
Higgs-boson couplings to the EW vector bosons and to fermions,
respectively, and do not introduce new couplings that are absent in
the SM. For the SM loop-induced couplings ($Hgg$, $H\gamma\gamma$, and
$HZ\gamma$) we assume that there is no contribution from new particles 
in the loop. 
For the relations between the scale factors and the Higgs-boson signal
strengths, we refer the reader to Ref.~\cite{Heinemeyer:2013tqa}. 

\begin{table}[t]
\centering
\begin{tabular}{c|c|c|rr}
\hline
& 68\% & 95\% & \multicolumn{2}{|c}{Correlations}
\\
\hline
$\kappa_V$ &
$1.02\pm 0.05$ & $[0.93,\, 1.11]$
& $1.00$ 
\\
$\kappa_f$ &
$0.97\pm 0.11$ & $[0.76,\, 1.20]$
& $0.22$ & $1.00$
\\
\hline
\end{tabular}
\caption{SM-like solution in the fit of $\kappa_V$ and $\kappa_f$ 
  to the Higgs-boson signal strengths.} 
\label{tab:kV_kf}
\end{table}

\begin{figure}[t]
  \centering
  \vspace{-1mm}
  \includegraphics[width=.24\textwidth]{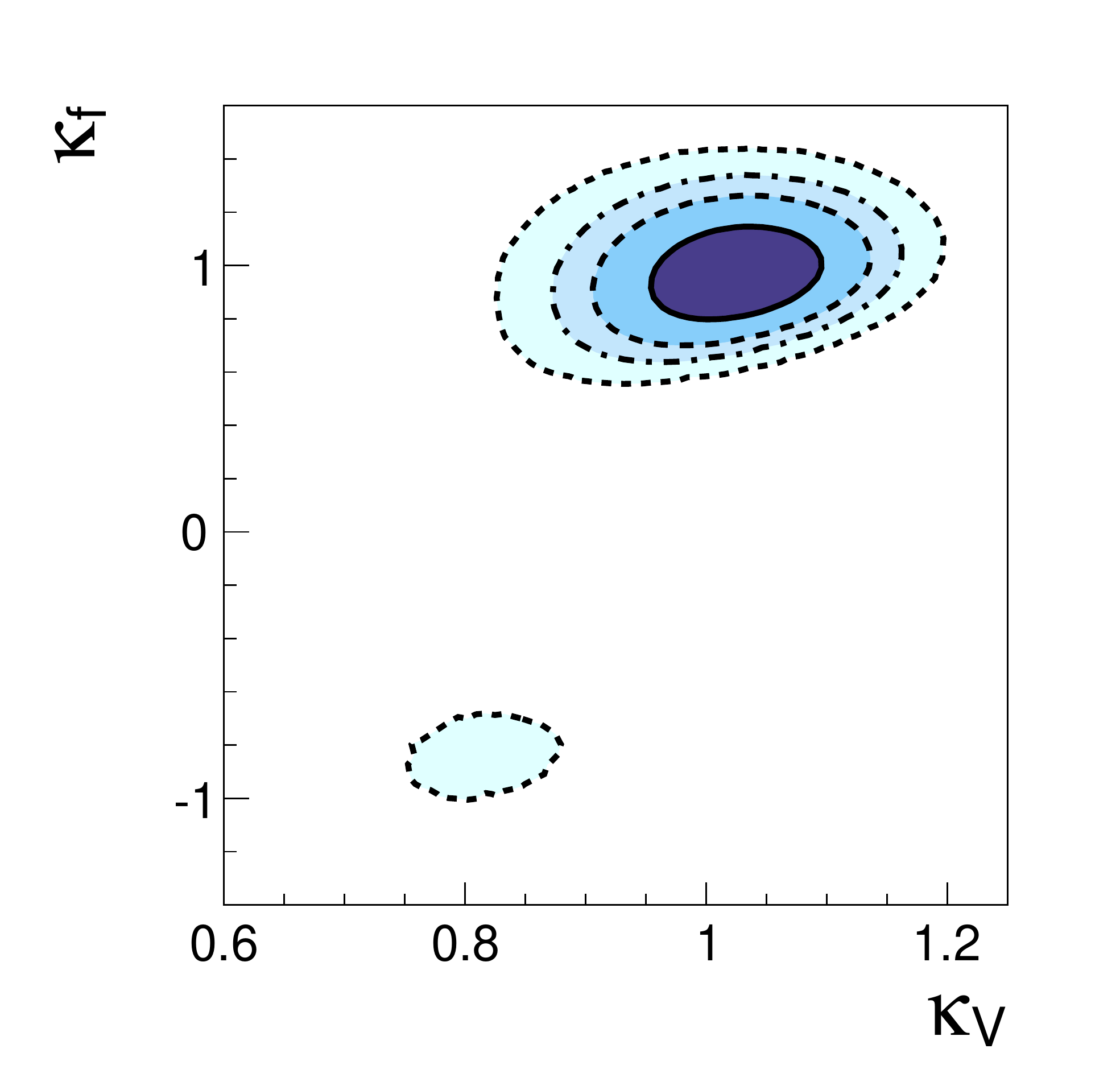}
  \hspace{-3mm}
  \includegraphics[width=.24\textwidth]{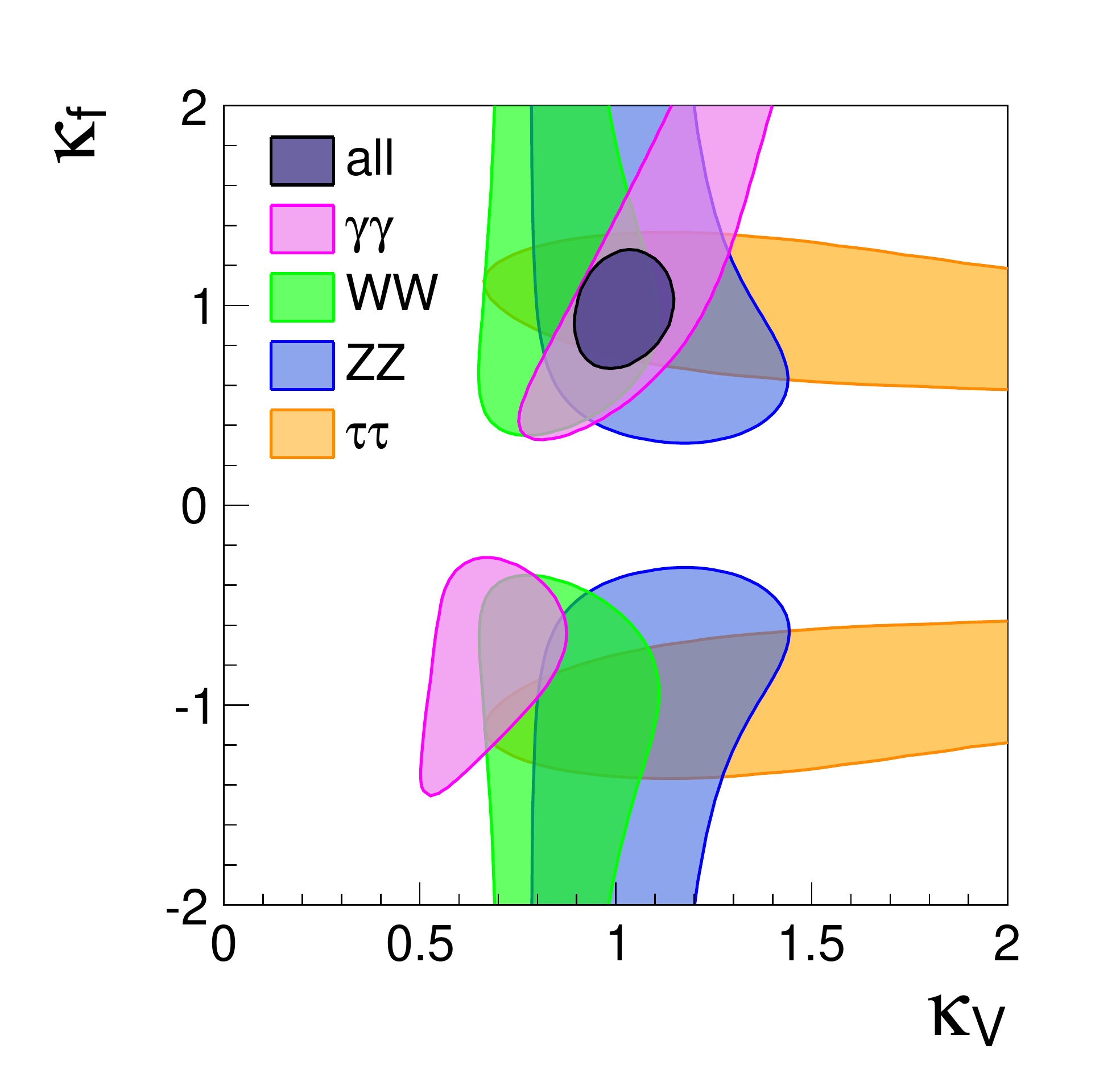}

  \vspace{-2mm}
  \caption{Left: Two-dimensional probability distributions for $\kappa_V$
    and $\kappa_f$ at 68\%, 95\%, 99\%, and 99.9\% (darker to lighter), 
    obtained from the fit to
    the Higgs-boson signal strengths. Right: Constraints from 
    individual channels at 95\%.}
  \label{fig:kV_kf}
\end{figure}

\begin{table}[t!]
\centering
\begin{tabular}{c|c|c|rr}
\hline
& 68\% & 95\% & \multicolumn{2}{|c}{Correlations}
\\
\hline
$\kappa_V$ &
$1.02\pm 0.02$ & $[0.99,\, 1.06]$
& $1.00$ 
\\
$\kappa_f$ &
$0.97\pm 0.11$ & $[0.77,\, 1.20]$
& $0.10$ & $1.00$
\\
\hline
\end{tabular}
\caption{Same as Table~\ref{tab:kV_kf}, 
  but considering both the Higgs-boson signal strengths and the EWPO.}
\label{tab:kV_kf-EW}
\end{table}

\begin{figure}[t!]
  \centering
  \vspace{-2mm}
  \includegraphics[width=.24\textwidth]{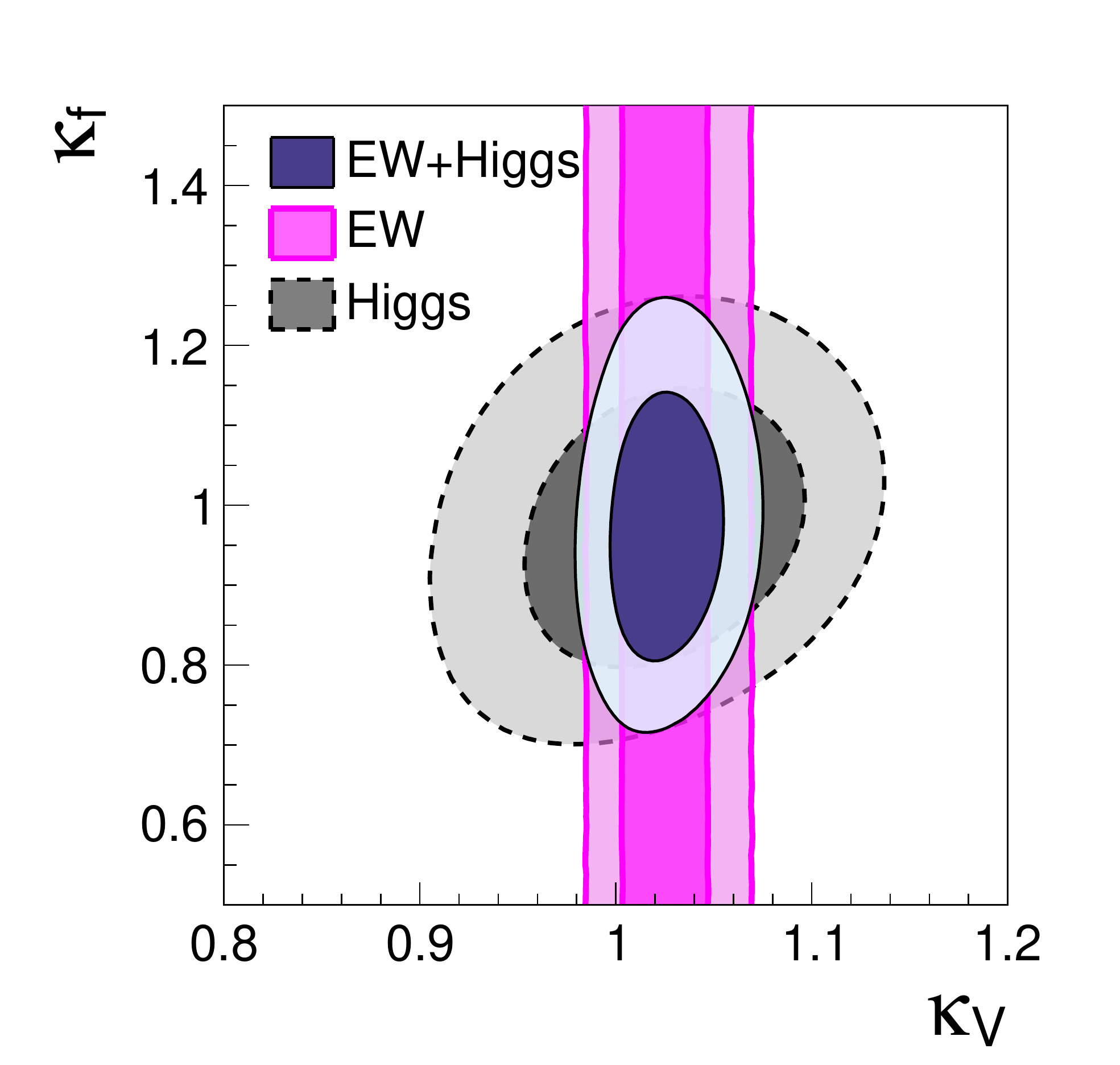}

  \vspace{-3mm}
  \caption{Two-dimensional probability distributions for $\kappa_V$
    and $\kappa_f$ at $68\%$ (the dark region) and $95\%$ (the light
    region), obtained from the fit to the Higgs-boson signal strengths and
    the EWPO.} 
  \label{fig:kV_kf-EW}
\end{figure}

In Table~\ref{tab:kV_kf} we summarize the fit results for $\kappa_V$
and $\kappa_f$ from the Higgs-boson signal strengths. Note that theoretical
predictions are symmetric under the exchange 
$\{\kappa_V,\ \kappa_f\} \leftrightarrow \{-\kappa_V,\ -\kappa_f\}$. 
In the left plot in Fig.~\ref{fig:kV_kf}, we present two-dimensional
probability distributions for $\kappa_V$ and $\kappa_f$ at 68\%, 95\%,
99\%, and 99.9\%, where only the parameter space with positive $\kappa_V$ is
presented. The region with negative $\kappa_f$ is disfavored in the fit. 
The right plot in Fig.~\ref{fig:kV_kf} shows constraints from the
individual decay channels. The constraints from $H\to b\bar{b}$ are
weaker than that from $H\to\tau^+\tau^-$ and are not presented for
simplicity. 
It is noted that because of the presence of flat directions in the
fit, the detailed shapes of the individual constraints depend on the
choice of the allowed ranges of the scale factors. 
We also consider constraints from the EWPO with the formulae in 
Eqs.~\eqref{eq:S} and \eqref{eq:T}, which are valid under the
assumptions given above Eq.~\eqref{eq:L_LightHiggs}. As shown in
Table~\ref{tab:kV_kf-EW} and Fig.~\ref{fig:kV_kf-EW}, the constraint
on $\kappa_V$ from the EWPO is stronger than that from the Higgs-boson 
signal strengths. 

\begin{figure}[t!]
  \centering
  \vspace{-1mm}
  \hspace*{-5mm}
  \begin{tabular}{lll}
  \includegraphics[width=29mm]{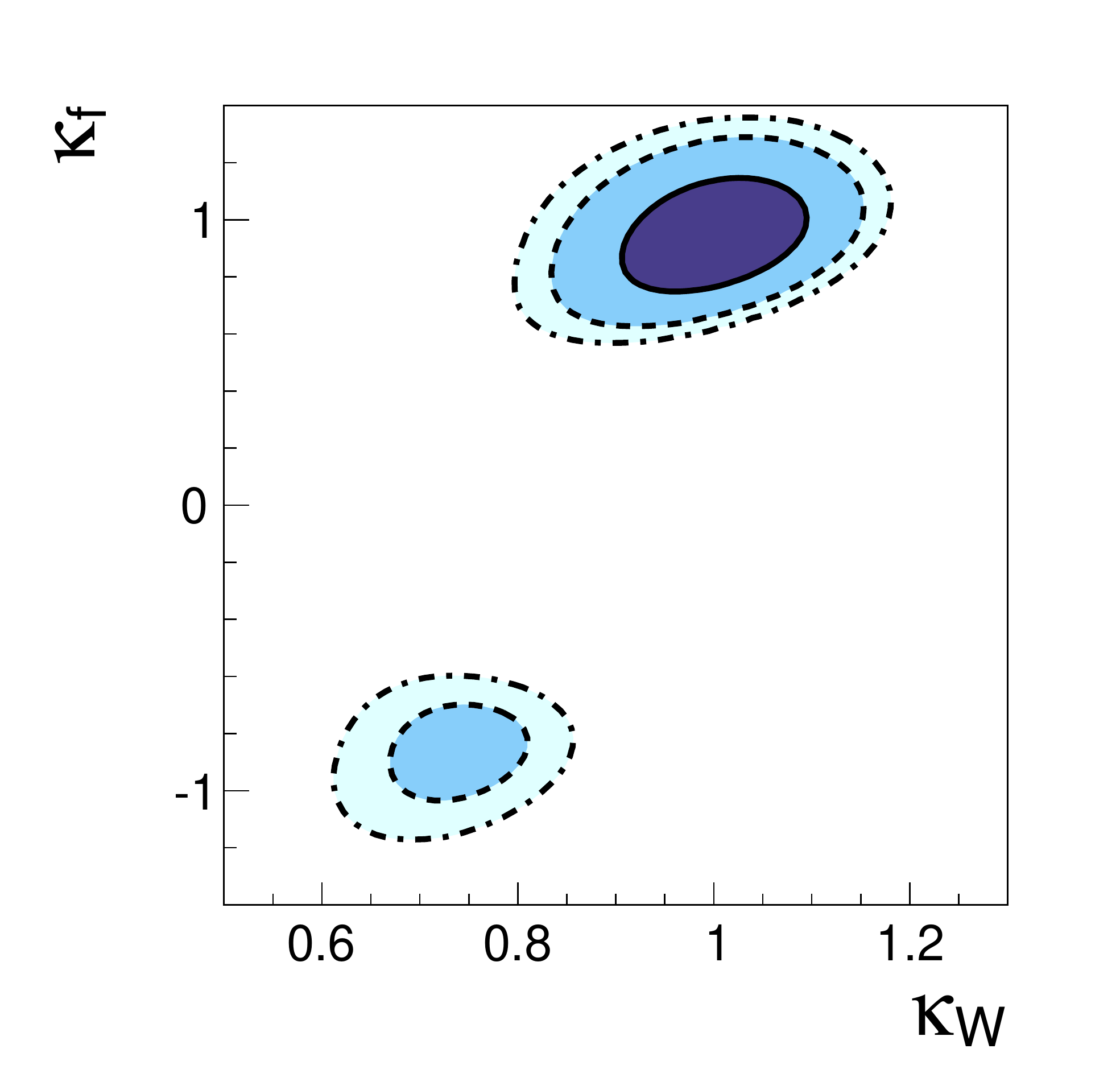}
  &\hspace{-7mm}
  \includegraphics[width=29mm]{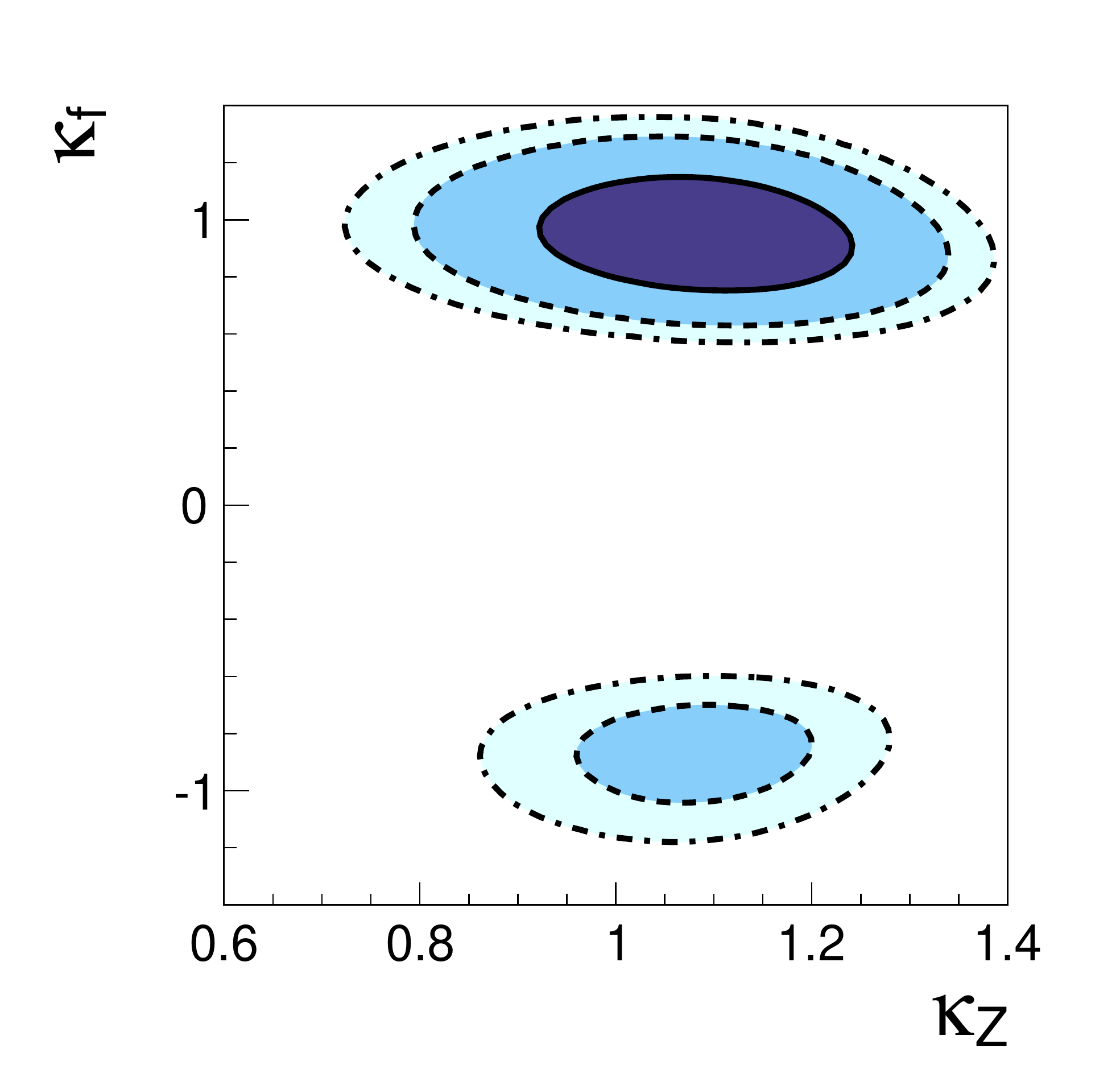}
  &\hspace{-7mm}
  \includegraphics[width=29mm]{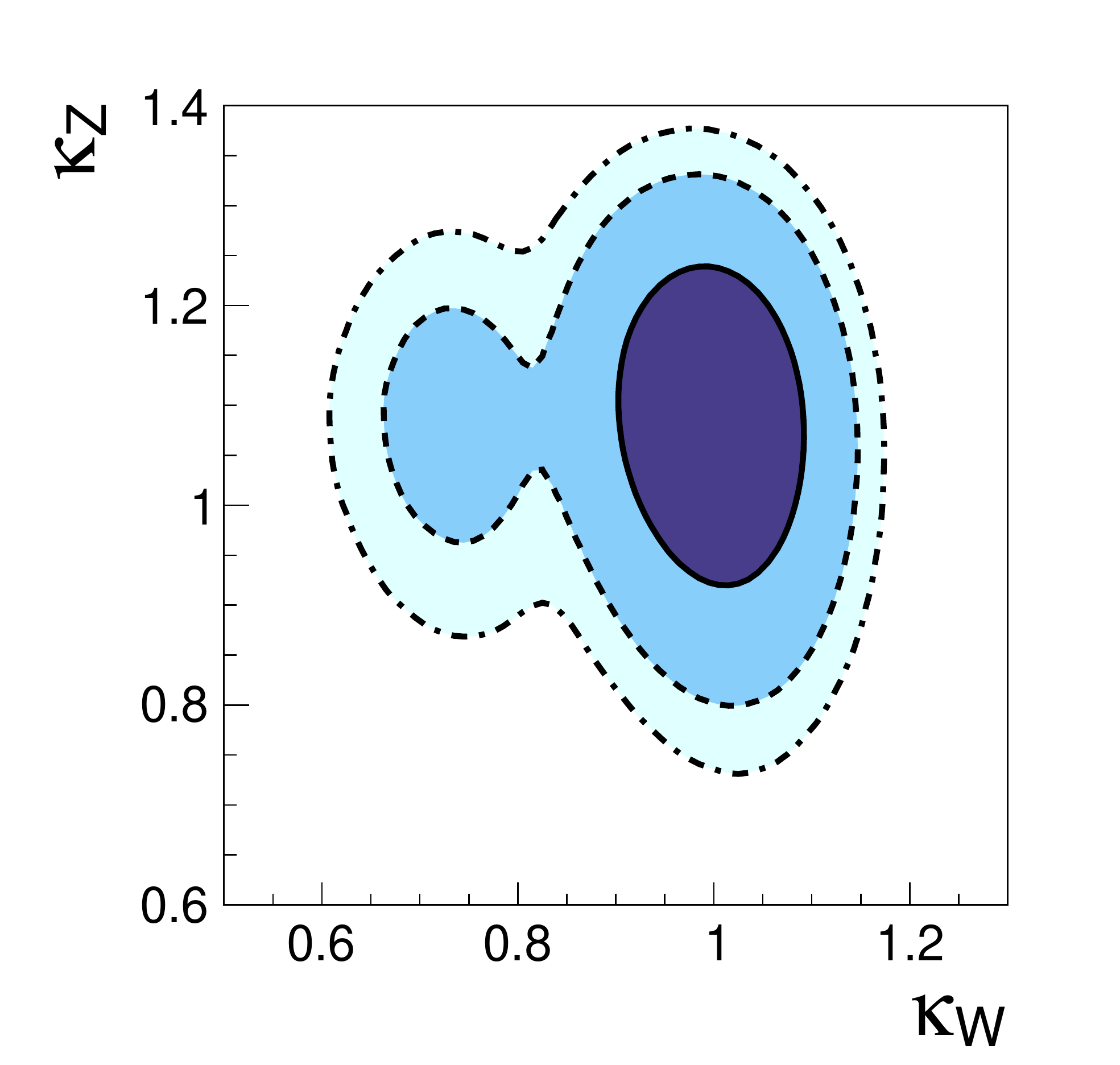}
  \end{tabular}
  
  \vspace{-2mm}
  \caption{Two-dimensional probability distributions for $\kappa_W$
    and $\kappa_f$ (left), for $\kappa_Z$ and $\kappa_f$
    (center), and for $\kappa_W$ and $\kappa_Z$ (right)
    at $68\%$, $95\%$, and $99\%$ (darker to lighter), obtained 
    from the fit to the Higgs-boson signal strengths.}
  \label{fig:kW_kZ_kf}
\end{figure}

Next we consider the case where the coupling to $W^+W^-$,
parameterized by $\kappa_W$, can differ from that to $ZZ$,
parameterized by $\kappa_Z$. 
Note that theoretical predictions are symmetric under the exchanges 
$\{\kappa_W,\ \kappa_f\} \leftrightarrow \{-\kappa_W,\ -\kappa_f\}$ 
and/or $\kappa_Z \leftrightarrow -\kappa_Z$, where $\kappa_Z$ can flip
the sign independent of $\kappa_W$, since the interference between the
$W$ and $Z$ contributions to the vector-boson fusion cross section is
negligible. Hence we consider only the parameter space where both
$\kappa_W$ and $\kappa_Z$ are positive. Here we do not consider the
EWPO, since $\kappa_W\neq\kappa_Z$ develops power divergences in the
oblique corrections. It means that the detailed information on UV
theory is necessary for calculating the oblique corrections. 
The fit results to the Higgs-boson signal strengths are summarized in 
Table~\ref{tab:kW_kZ_kf} and Fig.~\ref{fig:kW_kZ_kf}, which are
consistent with custodial symmetry. 

\begin{table}[t]
\setlength{\tabcolsep}{4pt}
\centering
\begin{tabular}{c|c|c|rrr}
\hline
& 68\% & 95\% & \multicolumn{3}{|c}{Correlations}
\\
\hline
$\kappa_W$ 
& $1.00\pm 0.06$ & $[0.88,\, 1.11]$
& $1.00$ 
\\
$\kappa_Z$ 
& $1.09\pm 0.10$ & $[0.88,\, 1.27]$
& $-0.12$ & $1.00$ 
\\
$\kappa_f$ 
& $0.94\pm 0.12$ & $[0.72,\, 1.18]$
& $0.35$ & $-0.16$ & $1.00$
\\
\hline
\end{tabular}
\caption{SM-like solution in the fit of $\kappa_W$, $\kappa_Z$, and
  $\kappa_f$ to the Higgs-boson signal strengths.} 
\label{tab:kW_kZ_kf}
\end{table}

\begin{figure}[t!]
  \centering
  \hspace*{-5mm}
  \begin{tabular}{lll}
  \includegraphics[width=29mm]{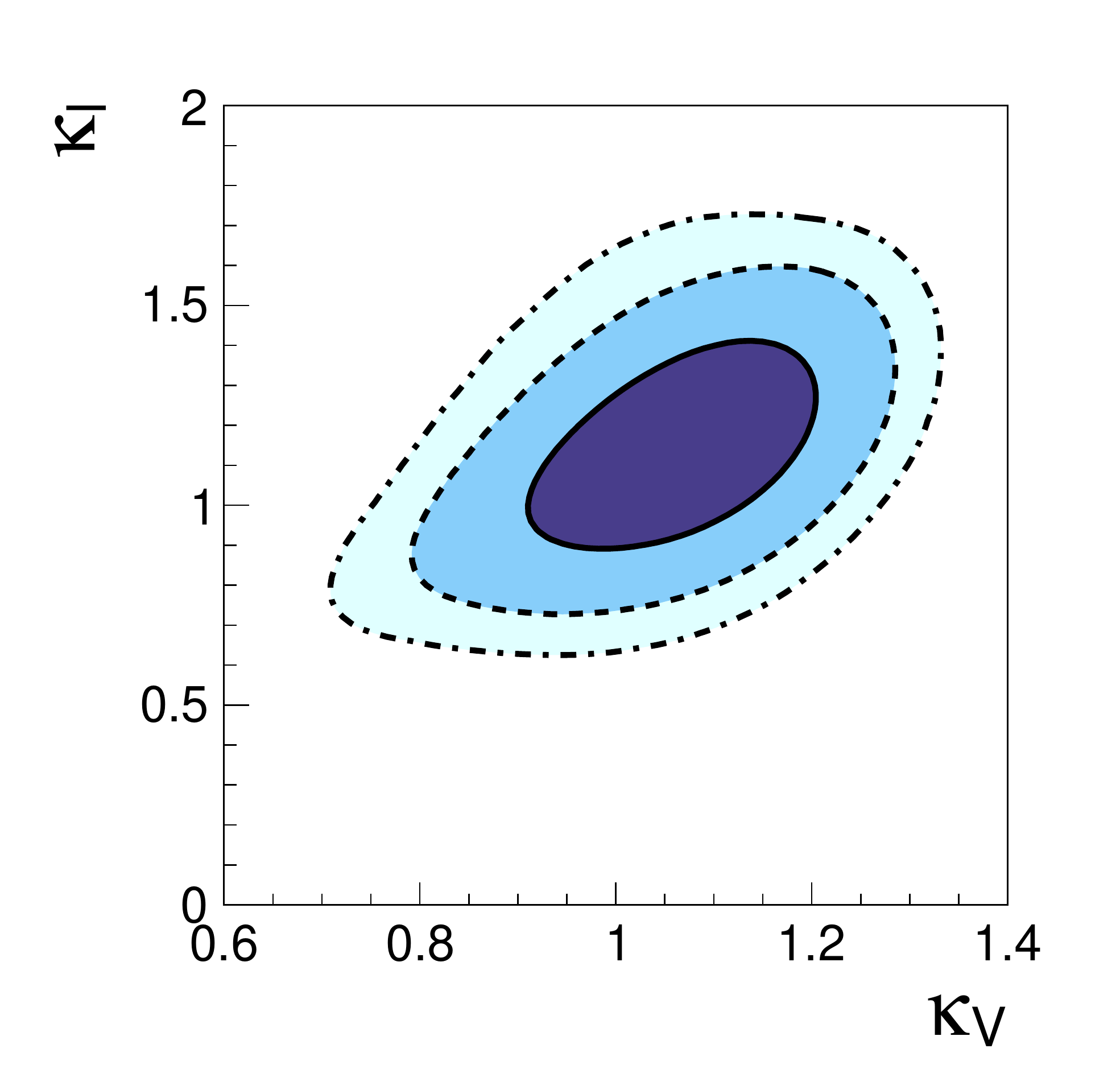}
  &\hspace{-7mm}
  \includegraphics[width=29mm]{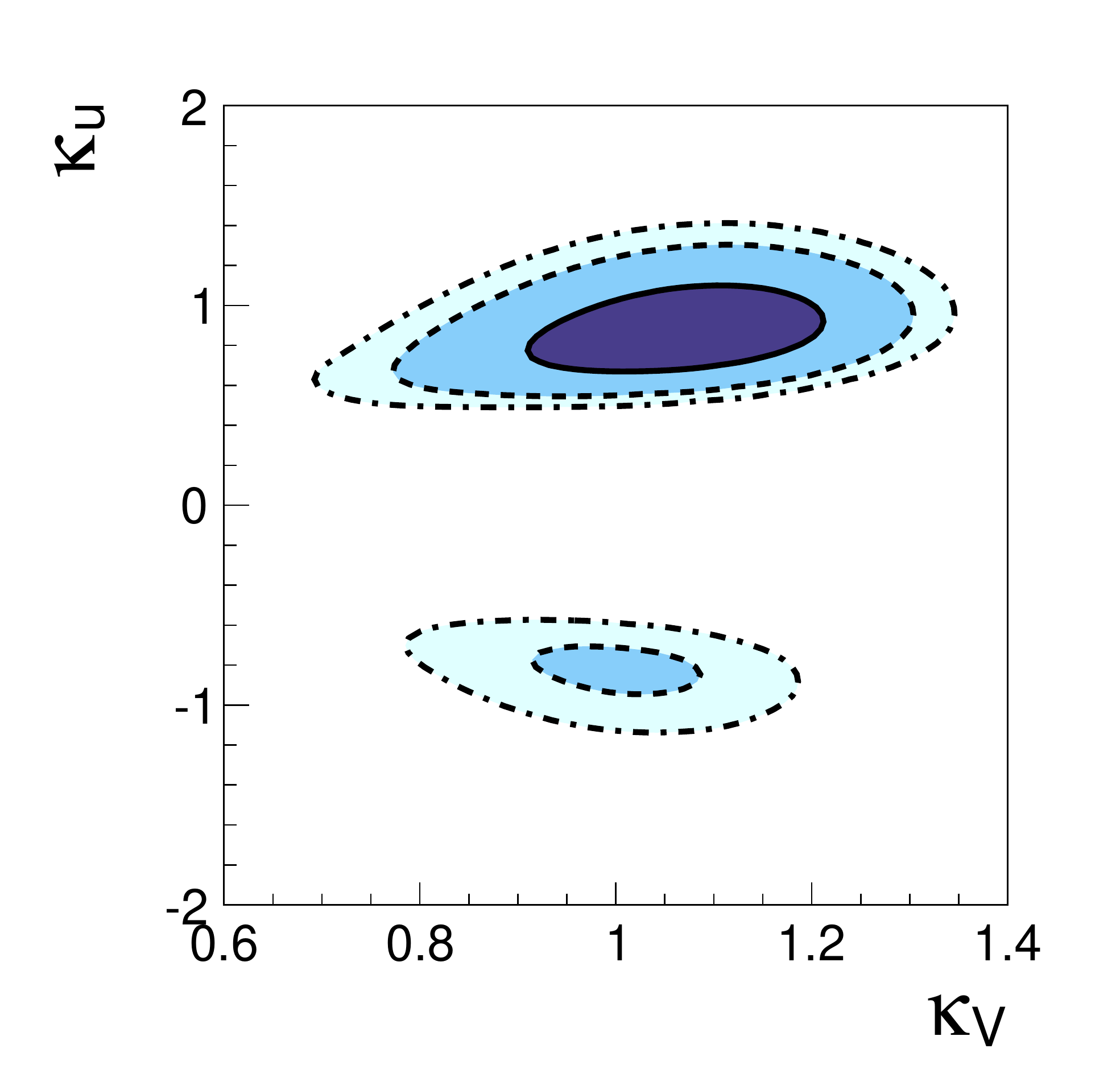}
  &\hspace{-7mm}
  \includegraphics[width=29mm]{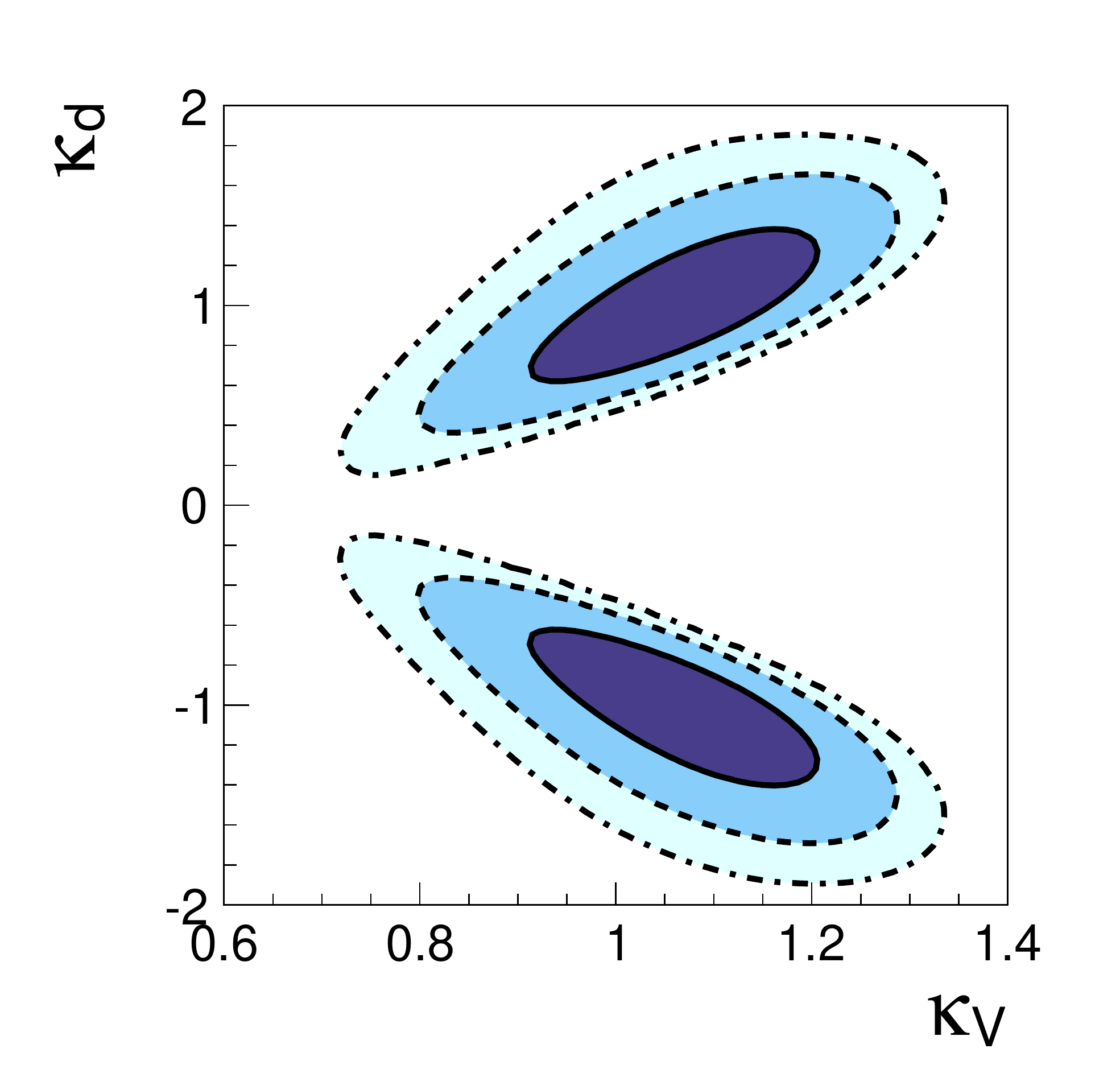}
  \\
  \includegraphics[width=29mm]{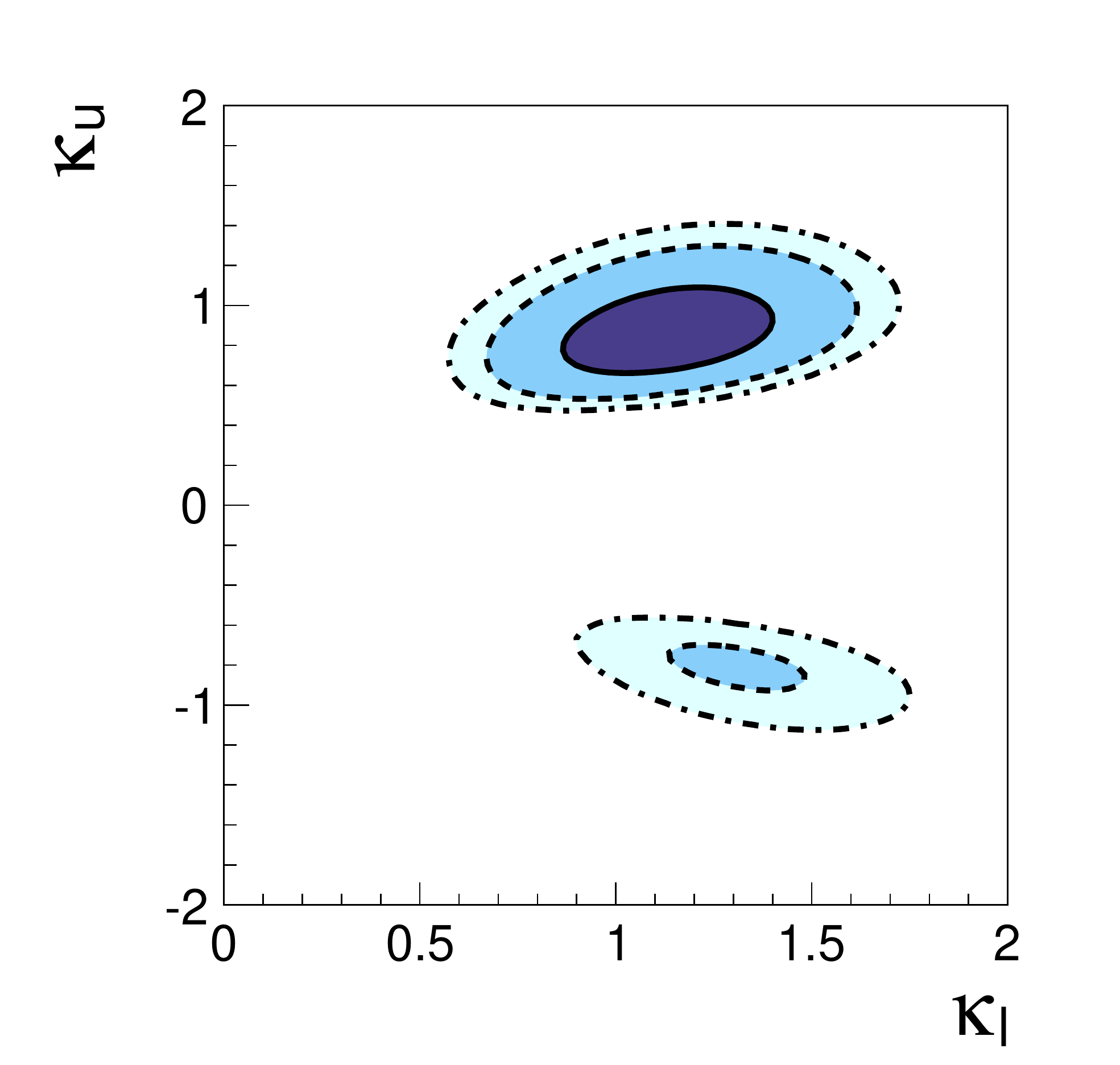}
  &\hspace{-7mm}
  \includegraphics[width=29mm]{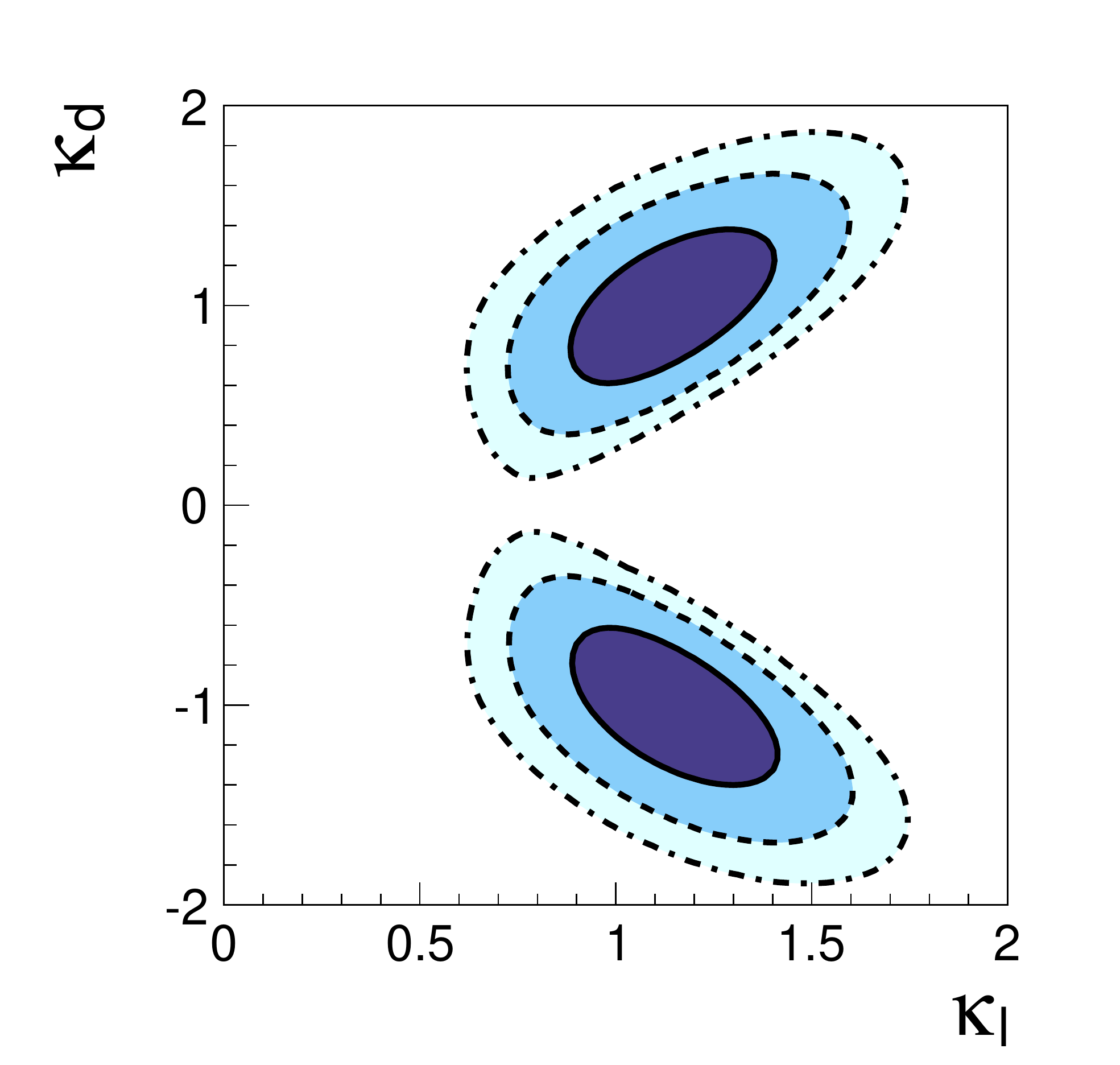}
  &\hspace{-7mm}
  \includegraphics[width=29mm]{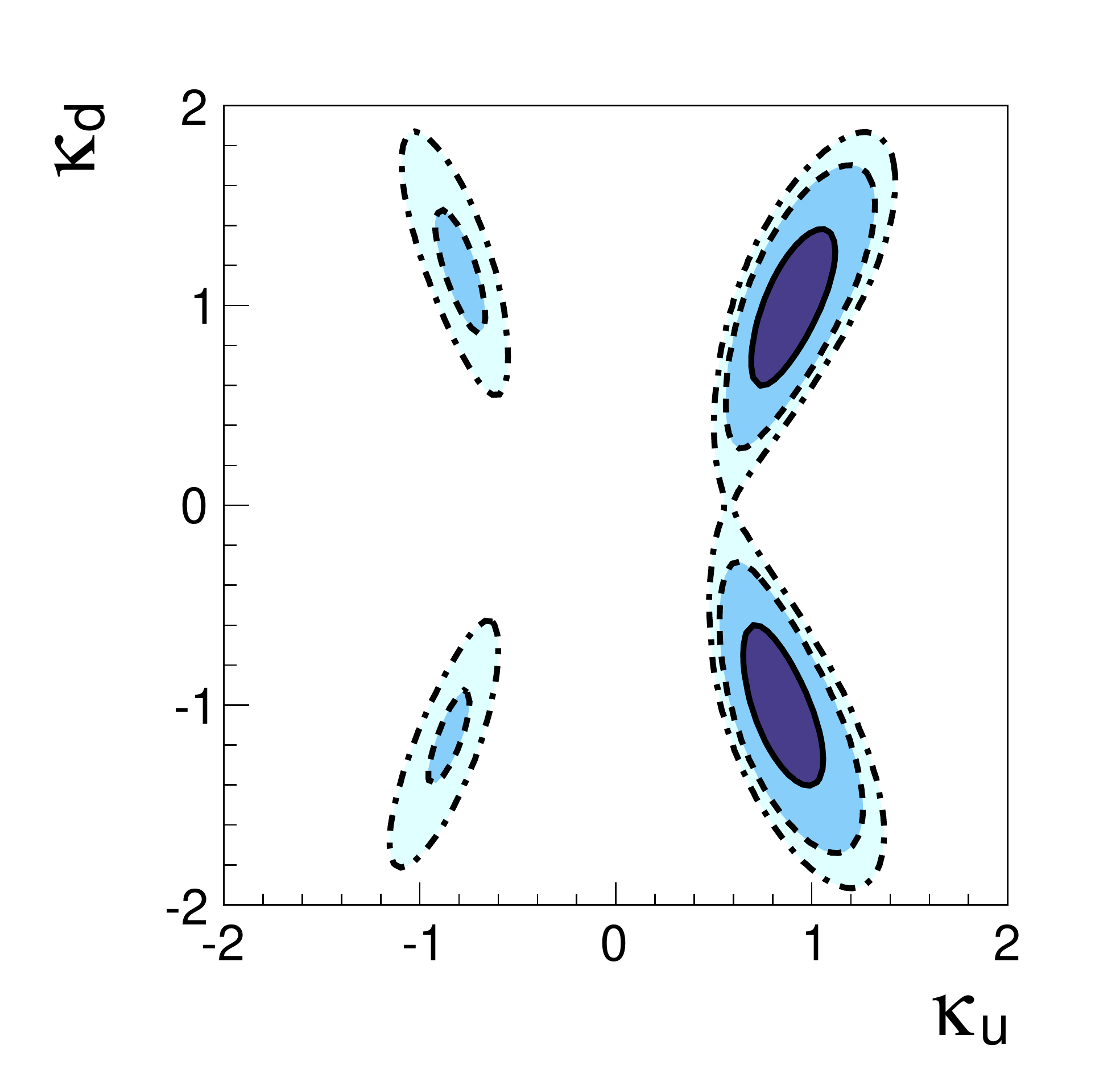}
  \end{tabular}
  
  \vspace{-2mm}
  \caption{Two-dimensional probability distributions
    for $\kappa_V$ and $\kappa_\ell$ (top-left), 
    for $\kappa_V$ and $\kappa_u$ (top-center), 
    for $\kappa_V$ and $\kappa_d$ (top-right), 
    for $\kappa_\ell$ and $\kappa_u$ (bottom-left), 
    for $\kappa_\ell$ and $\kappa_d$ (bottom-center), and 
    for $\kappa_u$ and $\kappa_d$ (bottom-right)
    at $68\%$, $95\%$, and $99\%$ (darker to lighter), obtained 
    from the fit to the Higgs-boson signal strengths.}
  \label{fig:kV_kl_ku_kd}
\end{figure}

\begin{figure}[t!]
  \centering
  \hspace*{-5mm}
  \begin{tabular}{lll}
  \includegraphics[width=29mm]{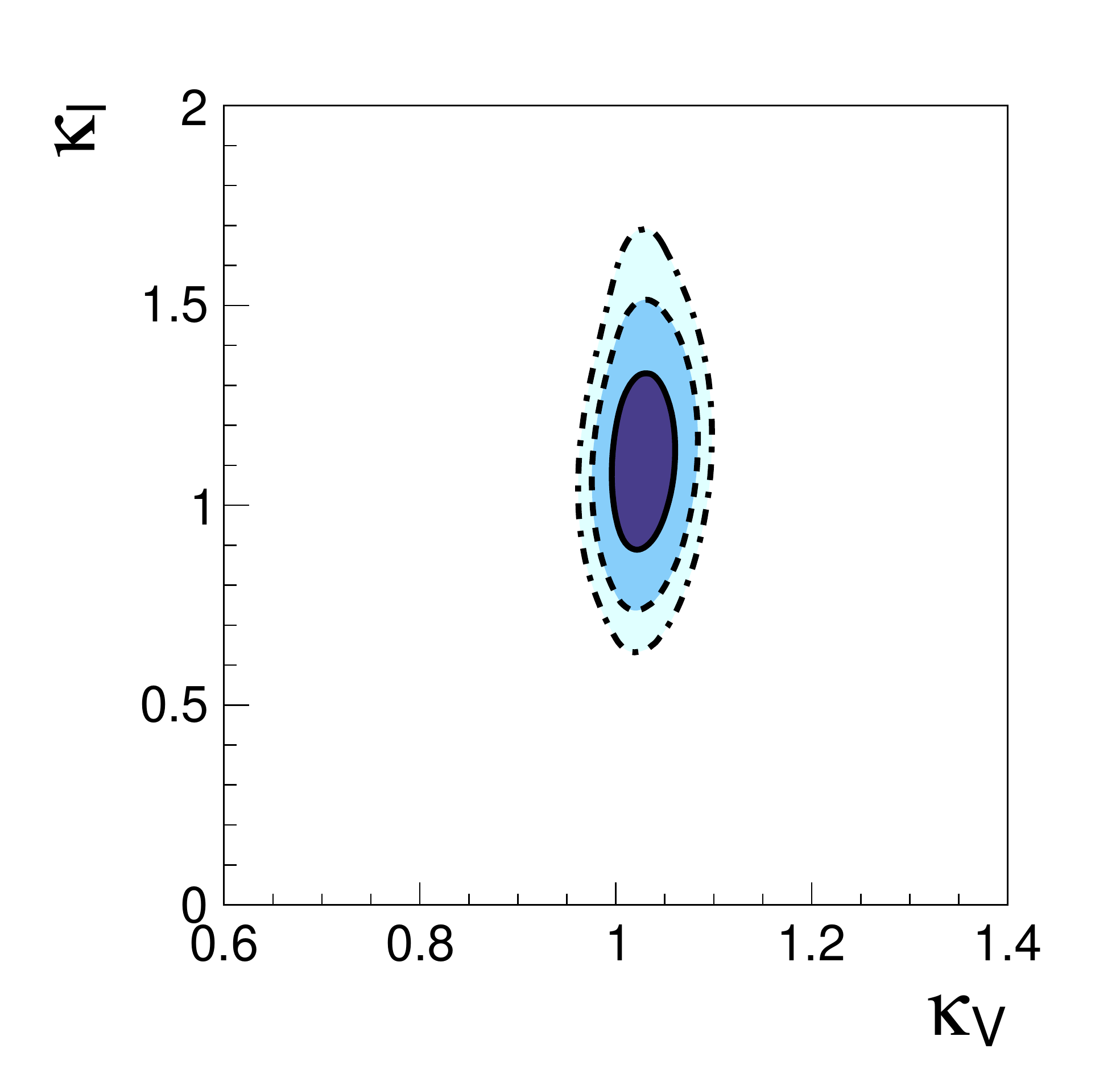}
  &\hspace{-7mm}
  \includegraphics[width=29mm]{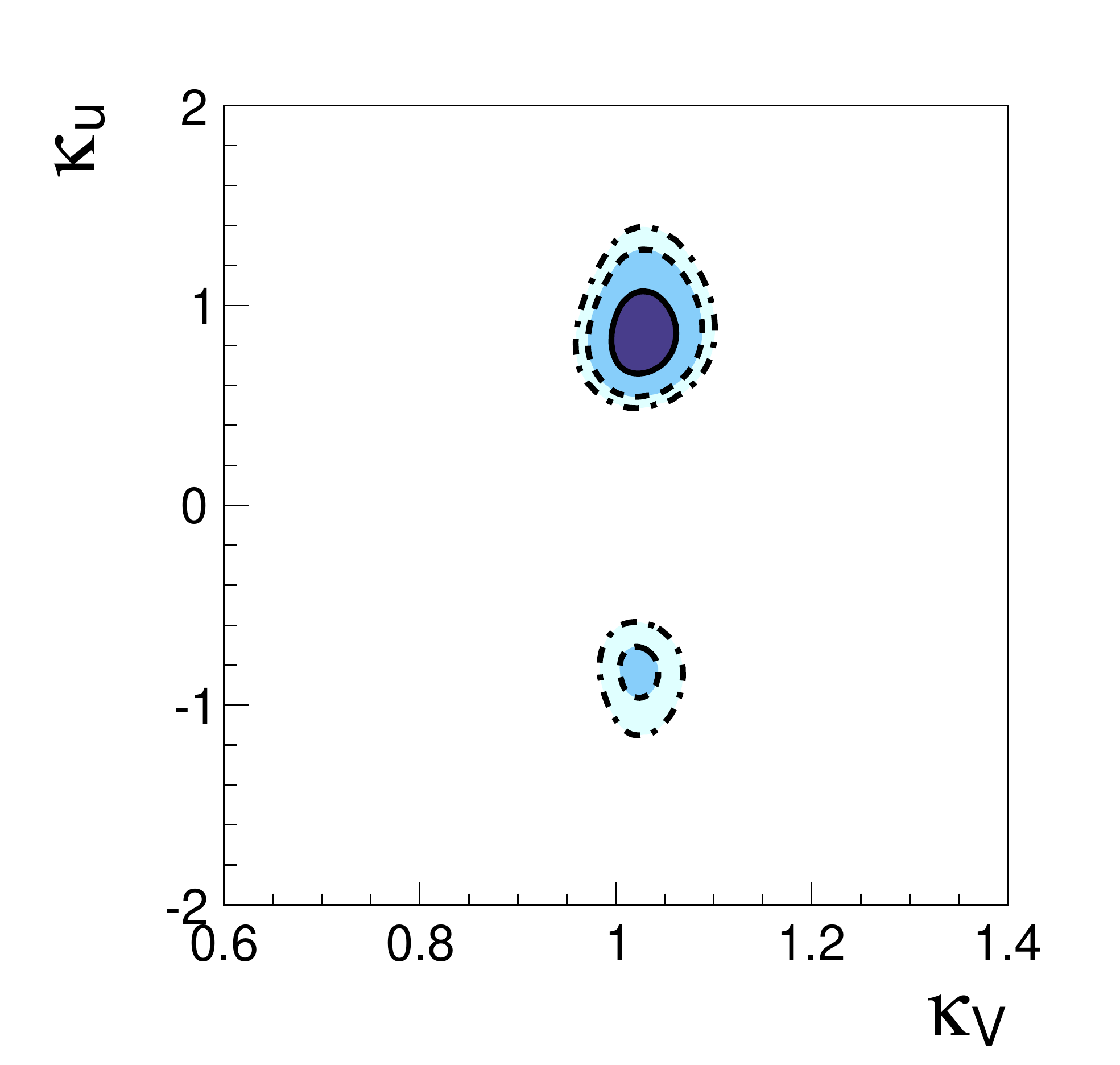}
  &\hspace{-7mm}
  \includegraphics[width=29mm]{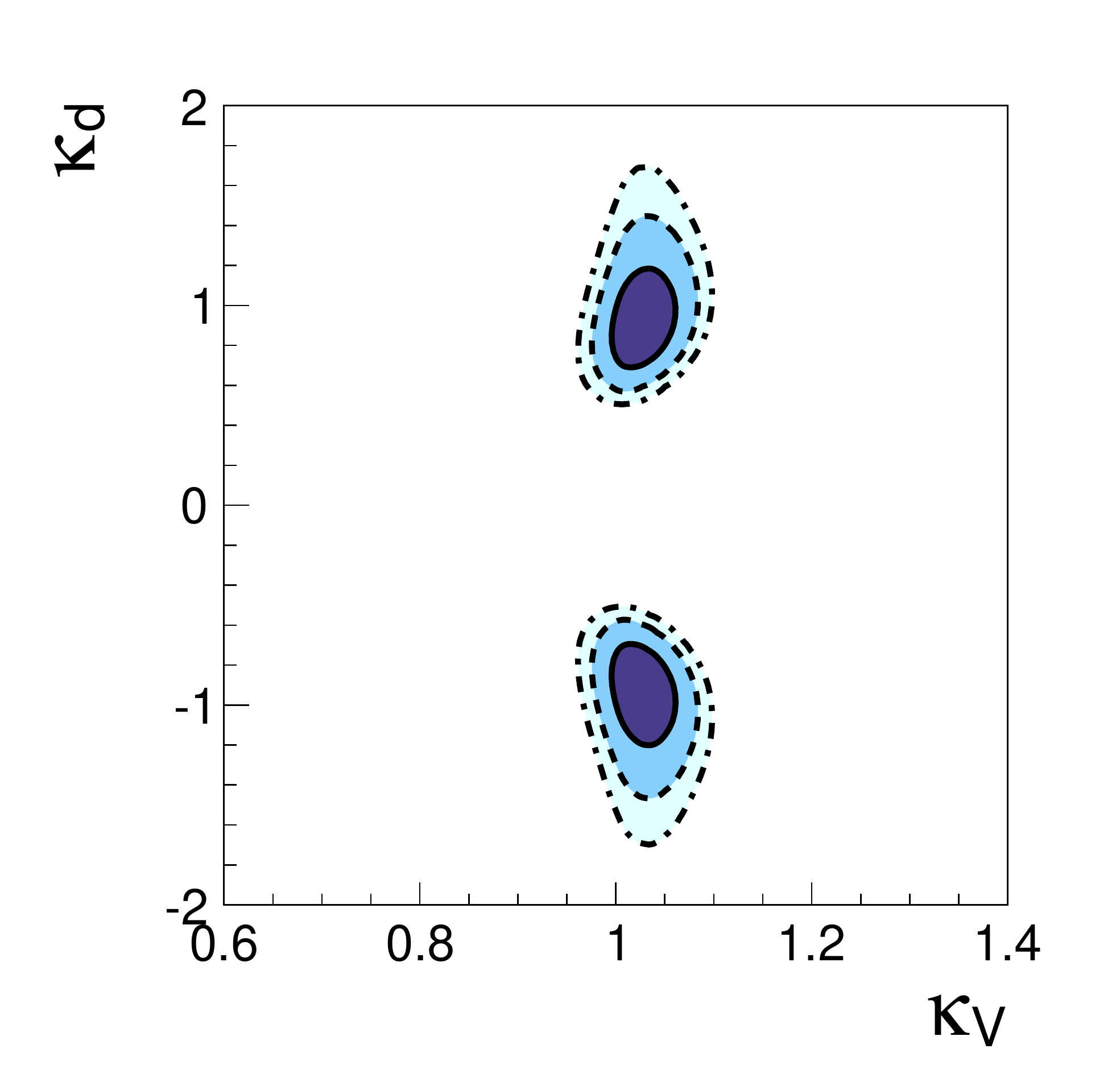}
  \\
  \includegraphics[width=29mm]{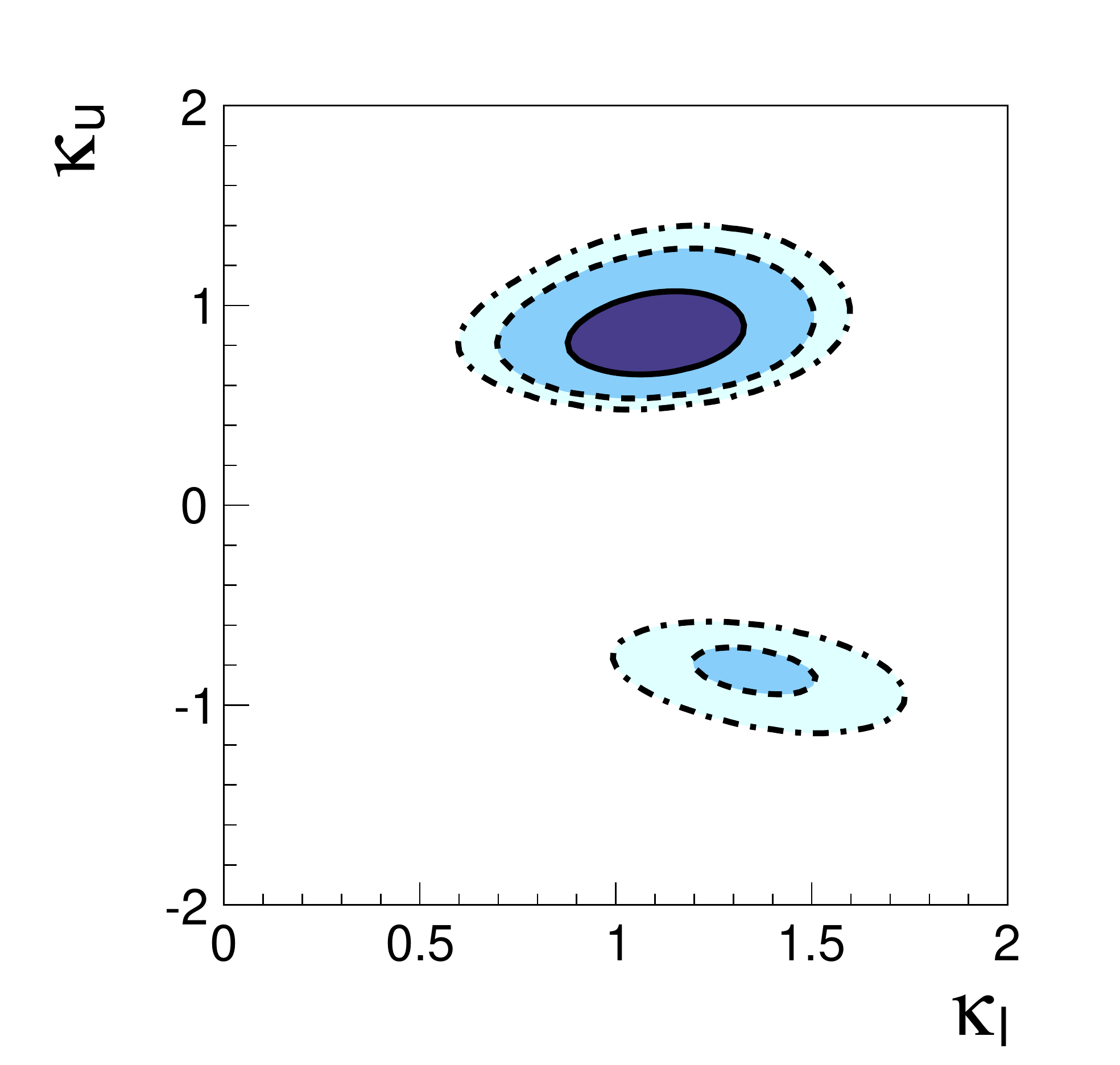}
  &\hspace{-7mm}
  \includegraphics[width=29mm]{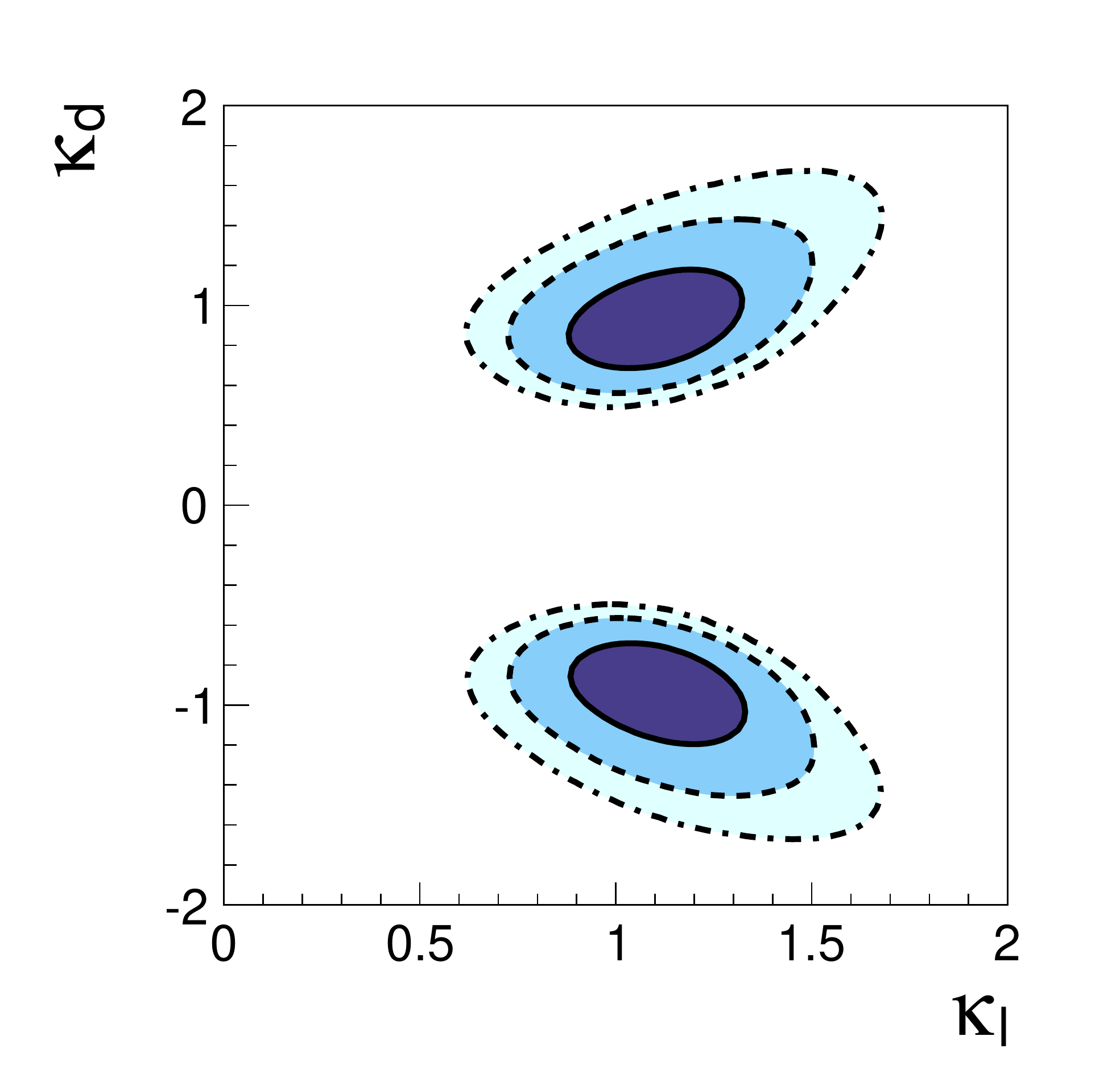}
  &\hspace{-7mm}
  \includegraphics[width=29mm]{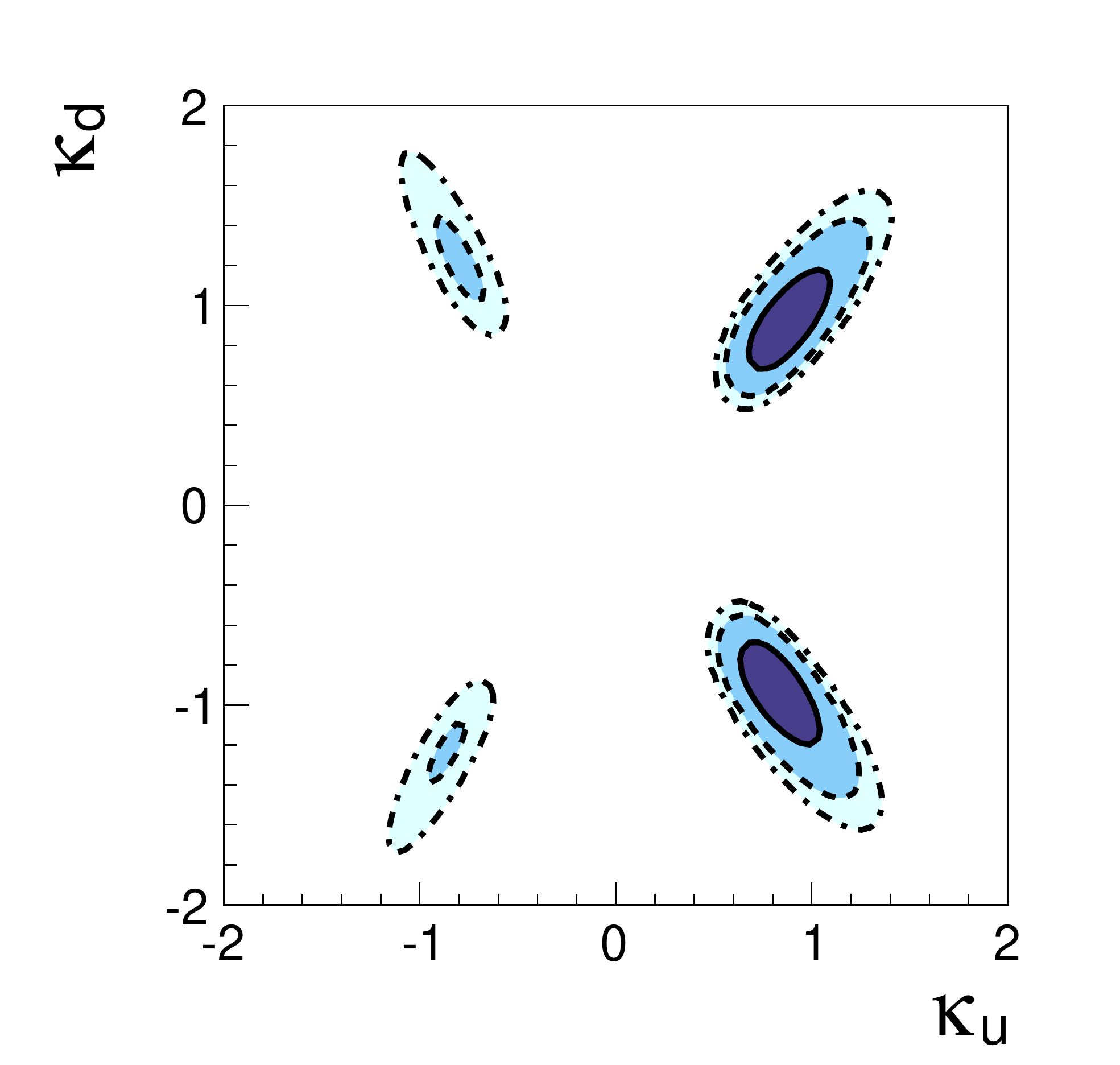}
  \end{tabular}
  
  \vspace{-2mm}
  \caption{Same as Fig.~\ref{fig:kV_kl_ku_kd}, 
  but considering both the Higgs-boson signal strengths and the EWPO.}
  \label{fig:kV_kl_ku_kd-EW}
  \vspace{-5mm}
\end{figure}

We also consider the case where the universality in the couplings to
the fermions is relaxed by introducing $\kappa_\ell$, $\kappa_u$ and
$\kappa_d$ for the couplings to the charged leptons, to the up-type
quarks, and to the down-type quarks. In this case, the Higgs-boson signal
strengths are symmetric under the exchanges 
$\kappa_\ell \leftrightarrow -\kappa_\ell$ and/or 
$\{\kappa_V,\ \kappa_u,\ \kappa_d\} \leftrightarrow
\{-\kappa_V,\ -\kappa_u,\ -\kappa_d\}$. Therefore, we consider only the
parameter space where both $\kappa_V$ and $\kappa_\ell$ are positive. 
The constraints on the scale factors from the Higgs-boson signal
strengths are presented in Table~\ref{tab:kV_kl_ku_kd} and
Fig.~\ref{fig:kV_kl_ku_kd}. By adding the EWPO to the fit, the
constraints become stronger as shown in Table~\ref{tab:kV_kl_ku_kd-EW}
and Fig.~\ref{fig:kV_kl_ku_kd-EW}.  

\begin{table}[t]
\setlength{\tabcolsep}{3pt}
\centering
\begin{tabular}{c|c|c|rrrr}
\hline
& 68\% & 95\% & \multicolumn{4}{|c}{Correlations}
\\
\hline
$\kappa_V$ &
$1.07\pm 0.09$ & $[0.87,\, 1.24]$
& $1.00$ 
\\
$\kappa_\ell$ &
$1.13\pm 0.17$ & $[0.80,\, 1.47]$
& $0.54$ & $1.00$
\\
$\kappa_u$ &
$0.89\pm 0.13$ & $[0.65,\, 1.18]$
& $0.37$ & $0.36$ & $1.00$
\\
$\kappa_d$ &
$1.01\pm 0.24$ & $[0.52,\, 1.51]$
& $0.79$ & $0.60$ & $0.75$ & $1.00$
\\
\hline
\end{tabular}
\caption{SM-like solution in the fit of $\kappa_V$, $\kappa_\ell$,
  $\kappa_u$, and $\kappa_d$ to the Higgs-boson signal strengths.} 
\label{tab:kV_kl_ku_kd}
\end{table}

\begin{table}[t!]
\setlength{\tabcolsep}{3pt}
\centering
\begin{tabular}{c|c|c|rrrr}
\hline
& 68\% & 95\% & \multicolumn{4}{|c}{Correlations}
\\
\hline
$\kappa_V$ &
$1.03\pm 0.02$ & $[0.99,\, 1.07]$
& $1.00$ 
\\
$\kappa_\ell$ &
$1.10\pm 0.14$ & $[0.82,\, 1.38]$
& $0.14$ & $1.00$
\\
$\kappa_u$ &
$0.88\pm 0.12$ & $[0.66,\, 1.15]$
& $0.09$ & $0.23$ & $1.00$
\\
$\kappa_d$ &
$0.92\pm 0.15$ & $[0.65,\, 1.26]$
& $0.28$ & $0.35$ & $0.81$ & $1.00$
\\
\hline
\end{tabular}
\caption{Same as Table~\ref{tab:kV_kl_ku_kd}, 
  but considering both the Higgs-boson signal strengths and the EWPO.}
\label{tab:kV_kl_ku_kd-EW}
\end{table}

\section{Summary}
\label{sec:summary}

We have updated the EW precision fits in the SM and beyond taking into
account the recent theoretical and experimental developments. 
The results of the SM fit are presented in Table~\ref{tab:SMfit}, 
while the constraints on the NP parameters (the oblique and epsilon
parameters, and the modified $Zb\bar{b}$ and $HVV$ couplings) are
summarized in Tables~\ref{tab:STUfit}-\ref{tab:HVV}. 
Furthermore, we have performed fits of the scale factors of the
Higgs-boson couplings to the Higgs-boson signal strengths and the EW
precision data as summarized in
Tables~\ref{tab:kV_kf}-\ref{tab:kV_kl_ku_kd-EW}.  
More detailed analyses and results will be presented in a future
publication~\cite{CFMPRS}.

\section*{Acknowledgments}

M.C. is associated to the Dipartimento di Matematica e Fisica,
Universit\`a di Roma Tre, and E.F. and L.S. are associated to the
Dipartimento di Fisica, Universit\`a di Roma ``La Sapienza''. 
We thank J.~de Blas and D. Ghosh for useful discussions and comments.
The research leading to these results has received funding from the
European Research Council under the European Union's Seventh Framework
Programme (FP/2007-2013) / grants n.~267985 and n.~279972. 
The work of L.R. is supported in part by the U.S. Department of Energy
under grant DE-FG02-13ER41942.




\nocite{*}
\bibliographystyle{elsarticle-num}
\bibliography{EWfit-ICHEP2014-v2}







\end{document}